\documentclass[12pt]{iopart}
\usepackage{iopams}
\usepackage[numbers]{natbib}
\usepackage{graphicx}
\usepackage[utf8]{inputenc}
\usepackage[T1]{fontenc}
\usepackage{dcolumn}
\usepackage{bm}
\usepackage{iopams} 
\usepackage{color}
\usepackage{soul}
\usepackage{url}
\usepackage[normalem]{ulem}
\expandafter\let\csname equation*\endcsname\relax
\expandafter\let\csname endequation*\endcsname\relax
\usepackage{amsmath}
\usepackage{siunitx}
\usepackage{aas_macros}

\usepackage[caption=false]{subfig}
\usepackage{tikz}

\tikzset{%
  base/.style = {rectangle, rounded corners, draw=black,
                  minimum width=0.45\columnwidth, minimum height=3.em,
                  text centered, font=\sffamily},
  inpu/.style = {base, fill=blue!15},
  conv/.style = {base, fill=red!15},
  dens/.style = {base, fill=green!15},
  outp/.style = {base, fill=yellow!15},
}

\usepackage{pgfplots}

\definecolor{pink}{RGB}{255, 20, 147}

\definecolor{amethyst}{rgb}{0.6, 0.4, 0.8}

\newcommand*{\unie}{European Gravitational Observatory (EGO), I-56021 Cascina, Pisa, Italy.}
\newcommand*{\uniz}{Scuola Normale Superiore (SNS), Piazza dei Cavalieri, 7 - 56126 Pisa, Italy.}
\newcommand*{\unia}{OzGrav, Swinburne University of Technology, Hawthorn, Melbourne, VIC 3122, Australia.}
\newcommand*{\unib}{Institute of Multi-messenger Astrophysics and Cosmology, Missouri University of Science and Technology, 1315 N.\ Pine St., Rolla, MO 65409, USA.}
\newcommand*{\MonashSPA}{School of Physics and Astronomy, Monash University, Vic 3800, Australia.}
\newcommand*{\OzGravMonash}{OzGrav: The ARC Centre of Excellence for Gravitational Wave Discovery, Clayton VIC 3800, Australia.}
\newcommand*{\unic}{Center for Interdisciplinary Exploration \& Research in Astrophysics (CIERA), Northwestern University, Evanston, IL 60208, USA.}
\newcommand*{\unid}{LIGO, California Institute of Technology, Pasadena, CA 91125, USA.}
\newcommand*{\unif}{SUPA, University of Glasgow, Glasgow G12 8QQ, United Kingdom.}
\newcommand*{\unig}{Max Planck Institute for Intelligent Systems, Max-Planck-Ring 4, 72076 T\"ubingen, Germany\\
Max Planck ETH Center for Learning Systems, Max-Planck-Ring 4, 72076 T\"ubingen, Germany}
\newcommand*{\unip}{Universit\'e de Paris,
CNRS, Astroparticule et Cosmologie, F-75013 Paris, France.}
\newcommand*{\unitorv}{Universit{\`a} di Roma Tor Vergata, I-00133 Roma, Italy.}
\newcommand*{\infntorv}{Istituto Nazionale di Fisica Nucleare, Sezione di Roma Tor Vergata, I-00133 Roma, Italy.}
\newcommand*{\cucal}{Columbia Astrophysics Laboratory, Columbia University in the City of New York, 550 W 120th St., New York, NY 10027, USA.}
\newcommand*{\cupd}{Department of Physics, Columbia University in the City of New York, 550 W 120th St., New York, NY 10027, US.A}
\newcommand*{\unih}{LIGO, Massachusetts Institute of Technology, Camrbidge, MA,  02139, USA.}
\newcommand*{\unii}{University of Minnesota, Minneapolis, MN 55455, USA.}
\newcommand*{\unipi}{Department of Physics, University of Pisa, Pisa, I-56127, Italy.}
\newcommand*{\infnpi}{Istituto Nazionale di Fisica Nucleare, Sezione di Pisa, Pisa, I-56127, Italy.}
\newcommand*{\camk}{Nicolaus Copernicus Astronomical Center, Polish Academy of Sciences, Bartycka 18, 00-716, Warsaw, Poland.}
\newcommand*{\uib}{Universitat de les Illes Balears, IAC3--IEEC, E-07122 Palma de Mallorca, Spain.}
\newcommand{\aeip}{Max Planck Institute for Gravitational Physics (Albert Einstein Institute), Am M\"uhlenberg 1, Potsdam 14476, Germany.}
\newcommand*{\mini}{School of Physics and Astronomy, University of Minnesota,
Minneapolis, Minnesota 55455, USA}

\def\apjl{Astrophys.\ J. Lett.}
\def\apjs{Astrophys.\ J. Supp.\ Ser.}

\def\aap{Astron.\ Astrophys.}

\usepackage[utf8]{inputenc}

\begin{document}

\title{Enhancing Gravitational-Wave Science with Machine Learning}

 
\author{Elena Cuoco} \ead{elena.cuoco@ego-gw.it} \address{\unie}\address{\uniz}\address{\infnpi}
\author{Jade Powell} \address{\unia}
\author{Marco Cavagli{\`a}} \address{\unib}
\author{Kendall Ackley}  \address{\MonashSPA}\address{\OzGravMonash}
\author{Micha{\l} Bejger} \address{\camk}
\author{Chayan Chatterjee} \address{\OzGravMonash} \address{Department of Physics, The University of Western Australia}
\author{Michael Coughlin}  \address{\unid}\address{\mini}
\author{Scott Coughlin}  \address{\unic}
\author{Paul Easter} \address{\MonashSPA}\address{\OzGravMonash}
\author{Reed Essick} \address{Kavli Institute for Cosmological Physics, University of Chicago, Chicago, IL, USA.}
\author{Hunter Gabbard}  \address{\unif}
\author{Timothy Gebhard}  \address{\unig}
\author{Shaon Ghosh}\address{Montclair State University, Montclair, NJ, USA.}
\author{Le\"ila Haegel}  \address{\unip}
\author{Alberto Iess} \address{\unitorv} \address{\infntorv}
\author{David Keitel} \address{\uib}
\author{Zsuzsa M\'arka} \address{\cucal}
\author{Szabolcs M\'arka} \address{\cupd}
\author{Filip Morawski} \address{\camk}
\author{Tri Nguyen}  \address{\unih}
\author{Rich Ormiston}  \address{\unii}
\author{Michael P{\"u}rrer} \address{\aeip}
\author{Massimiliano Razzano}  \address{\unipi} \address{\infnpi}
\author{Kai Staats}  \address{\unic}
\author{Gabriele Vajente} \address{\unid}
\author{Daniel Williams}  \address{\unif}

\date{\today}

\begin{abstract}
\noindent
Machine learning has emerged as a popular and powerful approach for solving problems in astrophysics. We review applications of machine learning techniques for the analysis of ground-based gravitational-wave detector data. Examples include techniques for improving the sensitivity of Advanced LIGO and Advanced Virgo gravitational-wave searches, methods for fast measurements of the astrophysical parameters of gravitational-wave sources, and algorithms for reduction and characterization of non-astrophysical detector noise. These applications demonstrate how machine learning techniques may be harnessed to enhance the science that is possible with current and future gravitational-wave detectors.
\end{abstract}

\maketitle

\section{Introduction}
\label{sec:intro}

In 2015, the Advanced LIGO (LIGO) \cite{aLIGO} and Advanced Virgo (Virgo) \cite{AdVirgo} Gravitational-Wave (GW) detectors made the first observation of a GW signal from a stellar-mass Compact Binary Coalescence (CBC) system \cite{Abbott:2016blz,Abbott:2016nmj}. Nine additional Binary Black Hole (BBH) mergers and one Binary Neutron Star (BNS) merger were observed by LIGO and Virgo during the first two advanced detector observing runs (O1/O2) \cite{PhysRevX.9.031040}. The six month-long O3a run (April 1st -- October 1st, 2019) and the recently completed O3b run have yielded tens of new BBH, BNS and Neutron Star-Black Hole (NSBH) detections \cite{Abbott_2020, LIGOScientific:2020stg} and detection candidates \cite{graceDB}. The current rate of detections is expected to increase in future observing runs, as LIGO and Virgo approach their design sensitivity and additional detectors such as KAGRA \cite{Somiya:2011np,Akutsu:2018axf} and LIGO-India \cite{ligo-india:2011} join the network of ground-based GW observatories. The improved sensitivity of the GW network will allow scientists to gain insights into the origins and astrophysical distributions of CBC GW sources \cite{LIGOScientific:2018jsj}, test general relativity \cite{LIGOScientific:2019fpa}, and measure cosmological parameters such as the Hubble constant \cite{Abbott:2019yzh}. It may also lead to the discovery of GW signals from new source types, such as core-collapse supernovae (CCSNe) \cite{Abbott:2019pxc} and magnetars \cite{Abbott:2019dxx}, or a stochastic GW background of cosmological or astrophysical origin \cite{LIGOScientific:2019vic}. 

Despite all the initial successes, the future of GW astronomy is facing many challenges. Processing and analyzing the increased rate of detections in future observing runs will require researchers to streamline current search pipelines. Refined astrophysical investigations of GW sources and tests of the fundamental nature of gravity will require precise reconstructions of GW signals and accurate estimates of their statistical and systematic errors. Identification and mitigation of instrumental and environmental data artifacts will require the development of fast and efficient methods for detector and signal characterization.

Machine Learning (ML) algorithms are novel methods for tackling these issues.
The LIGO and Virgo Scientific Collaborations run searches for modeled and unmodeled astrophysical transient signals, as well as searches for continuous GWs from isolated compact objects and searches for a stochastic background of GWs of cosmological or astrophysical origin. These searches rely on different techniques, such as matched filtering \cite{wienerbook,PhysRevD.44.3819}, time-coincident detection of coherent excess power between multiple detectors \cite{2016PhRvD..93d2004K, 0264-9381-25-11-114029}, and cross-correlation methods \cite{PhysRevD.59.102001}. Because of the effectiveness of ML algorithms in identifying patterns in data, ML techniques may be harnessed to make all these searches more sensitive and robust. Applications of ML algorithms to GW searches range from building automated data analysis methods for low-latency pipelines to distinguishing terrestrial noise from astrophysical signals and improving the reach of searches, as well as the statistical significance of detections. 

In this paper, we review the ML-based techniques that have been developed by LIGO and Virgo scientists to improve the analysis of GW data. In recent years, ML has gained popularity among LIGO and Virgo researchers thanks to advances in detection and classification of noise transients \cite{powell:15, powell:16,PhysRevD.95.104059, PhysRevD.97.101501, 2013PhRvD..88f2003B,Pinto2013,Lightman_2006,2018razzano,Huerta:2019rtg,Cavaglia:2020qzp}, searches for CBC systems \cite{Kim_2015, Kim_2020, PhysRevLett.120.141103, 2018PhLB..778...64G}, parameter estimation of transient signals \cite{2012MNRAS.421..169G, 2019arXiv190301998S, 2018PhLB..778...64G,1909.06296,PhysRevLett.124.041102}, noise removal \cite{VaHu2019} and citizen science projects \cite{gravityspy}. We also examine the potential of ML to improve GW science as current detectors approach their design sensitivity and other detectors join the GW network of ground-based observatories. 

The structure of the paper is as follows. In Section \ref{sec:gw_data_glitch} we review the ML algorithms which have been developed to improve the quality of LIGO-Virgo data. In Section~\ref{sec:wf_model} we review ML techniques which aim at improving the modeling of GW signals. In Section \ref{sec:searches} we describe how ML can be used to improve the sensitivity of GW searches. In Section \ref{sec:pe} we review methods for parameter estimation of GW signals and source population inference. Conclusions are given in Section \ref{sec:conclusion}.

\section{Algorithms for gravitational-wave data quality improvement}
\label{sec:gw_data_glitch}

The output of a GW detector is a temporal series of the detector strain, $h(t)$. The sensitivity of an ideal detector is determined by the physics inherent to the instrumental design and is limited by fundamental, irreducible noise sources, such as quantum noise of the laser light, thermal noise of the mirror coatings and optic suspensions, or seismic noise \cite{aLIGO,AdVirgo}. The sensitivity of a real-world detector is also limited by the presence of technical noise sources of different origins, related for example to the feedback control systems that are needed to maintain the systems in operation, or by instrumental and environmental disturbances. Often these noise sources are non-stationary, i.e., their statistical properties vary over short or long time scales. In some instances, noise sources might be stationary at their origins, but couple to the detector strain in a nonlinear way.

The non-stationary, non-Gaussian nature of the data and the presence of noise artifacts 
may impact data quality or detector performance, and increase the false alarm rate of searches. Short-lived environmental and instrumental noise disturbances, known as detector ``glitches,'' may also affect low-latency detection and parameter estimation of astrophysical transient signals, as evidenced by the GW170817 BNS detection \cite{PhysRevX.9.031040}. These non-stationary and nonlinear transients, as well as continuous noise signals at given frequencies in the form of spectral lines, are two examples of major factors affecting the performance of GW searches \cite{AbEA2016g}. 

Astrophysical signals have typical amplitudes comparable to the detector background noise. Therefore, characterization and reduction of detector noise is essential to GW searches. Some of the techniques for the identification and mitigation of LIGO-Virgo data quality are described in Ref.~\cite{Abbott_2020}. 

\begin{figure}[t]
    \centering
    \includegraphics[width=15cm]{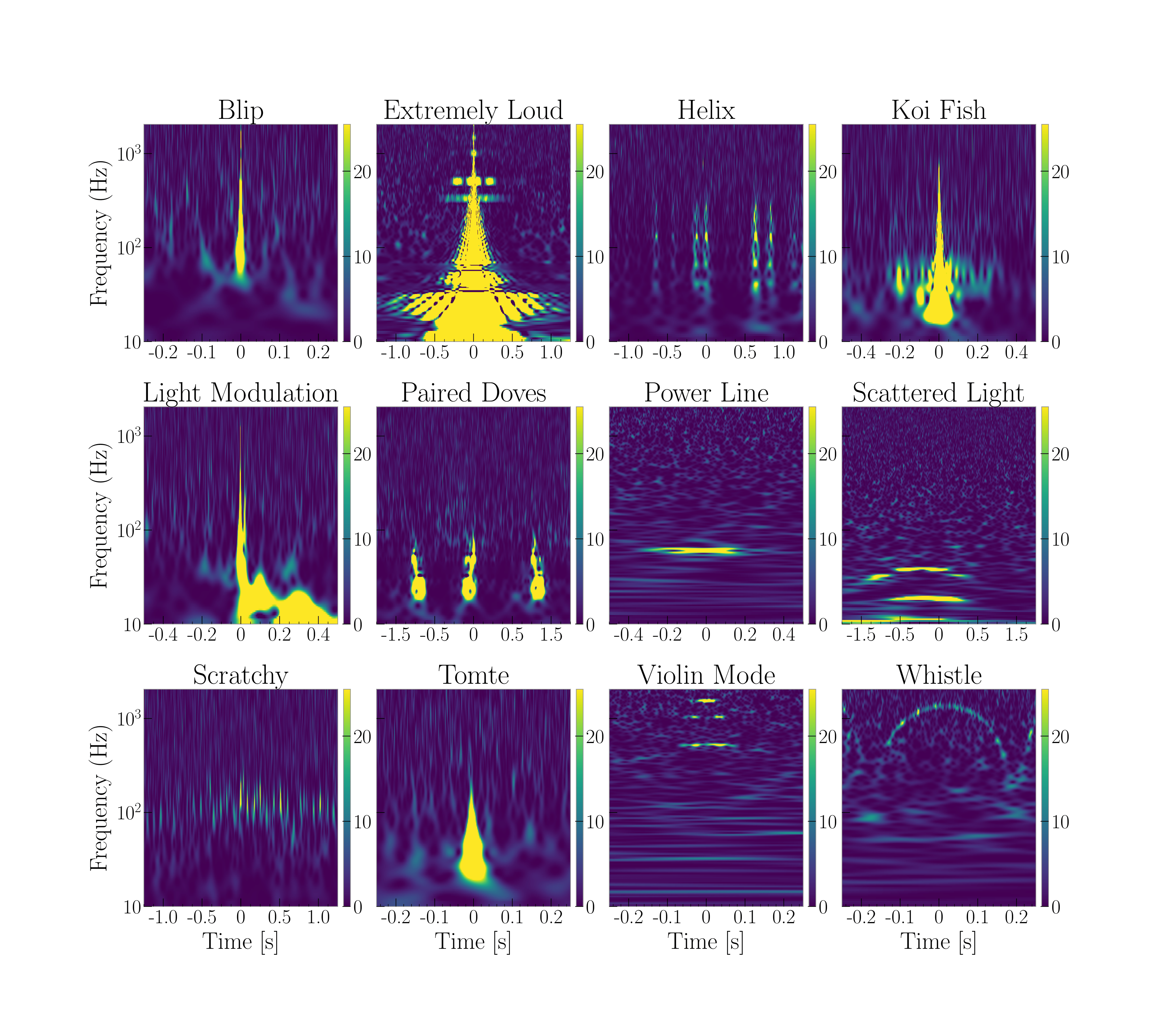}
    \caption{Time-frequency representations of different types of glitches occurring in GW data. (Examples from the GravitySpy project \cite{BAHAADINI2018172}). ML algorithms can help identify the origin of these glitches and increase the sensitivity of GW transient searches.}
    \label{fig:glitches}
\end{figure}

\subsection{Machine learning for $h(t)$ glitch characterization and classification}

The first step in the characterization of detector transient noise is to distinguish the glitches from potential astrophysical signals and then classify them into different families. This task can be tackled by extracting features from each glitch time series and mapping these features to the target glitch types. However, glitches often exhibit a complex temporal and frequency evolution that can make their characterization with a fixed number of features very difficult. Moreover, the increasing sensitivity of the  detectors may lead to a larger number of glitch morphologies. ML can help solve the problem of glitch classification. Earlier works on a variety of unsupervised ML methods \cite{Powell:2015ona,Powell:2016rkl} and neural networks \citep{2017Mukund} have shown that ML algorithms can be very effective. 

Deep Convolutional Neural Networks \cite{Yamashita2018,Russakovsky2015} (CNNs) are extremely promising for glitch classification based on time-frequency representations. 
CNNs are designed to extract features from 2D matrices, such as images, and use these features for classification purposes. Feeding time-frequency transforms, Omega Scans \cite{rollinsthesis}, and Q-transforms \cite{2004CQGra..21S1809C} to a CNN-based deep network is an effective approach to glitch classification. Ref.~\cite{2018razzano} implements an image-based detection and classification pipeline built on 2D CNN layers. Tests with GW simulated signals show that this method provides a $\sim 99$\% accuracy in classification and differentiation of glitches from chirp-like signals. CNNs typically provide higher accuracy of distinguishing glitches with similar morphology with respect to other ML approaches.

Along the same lines, Ref.~\cite{8553393} uses a Wavelet Detection Filter \cite{WDF-GRB} to extract the features from the input data set. The algorithm performs glitch classification with a boosted gradient method \cite{XGBoost} which could be suitable for real-time analysis. 

One critical step in supervised learning approaches is the availability of labeled glitch samples. The citizen science project GravitySpy \cite{gravityspy} addresses this issue. Based on the Zooniverse platform, GravitySpy leverages the advantages of citizen science and ML to design a socio-computational system to analyze and characterize transients in GW data. The citizen scientists classify images of glitches such as those shown in Figure \ref{fig:glitches}. The glitch categories resulting from this manual classification process are then used as labels for supervised ML approaches \cite{BAHAADINI2018172}. In addition, as the citizen scientists identify new categories of transients, the ML algorithms are re-trained to take into account the new categories.

\subsection{Glitch characterization and classification with auxiliary channels}

\begin{figure}[t]
    \centering
    \includegraphics[width=11cm]{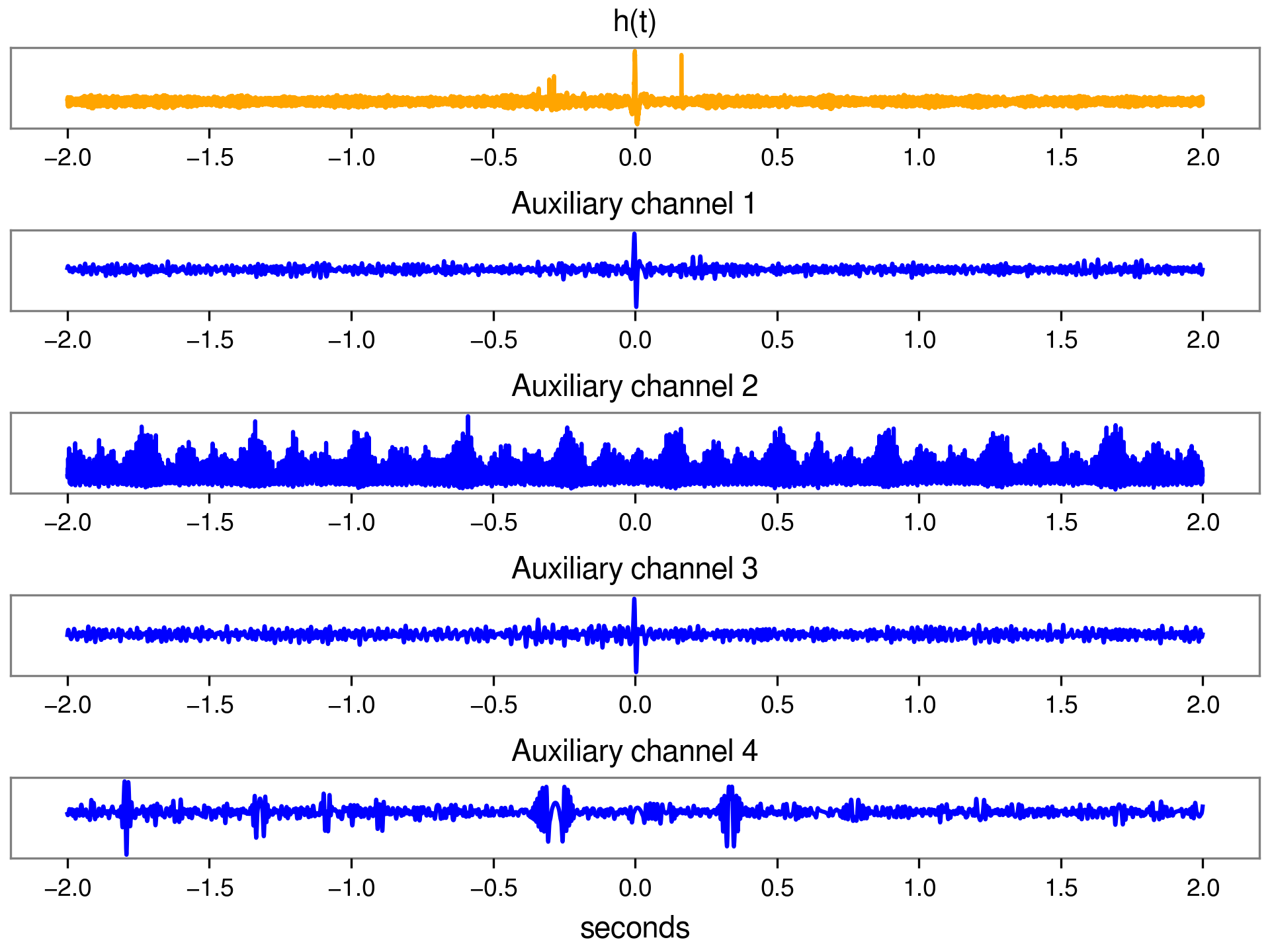}
    \caption{An example of how auxiliary channels can help determine the non-astrophysical nature of detector triggers. The top time series is $h(t)$. The other time series are from detector auxiliary monitoring different sources which are not sensitive to GWs. The spike in the strain time series at $t=0$ occurs also in auxiliary channel 1 and 3. Therefore, it can be concluded that the trigger is not of astrophysical origin. }
    \label{fig:channels}
\end{figure}

The LIGO and Virgo detectors record data streams from a large number of subsystems controlling different aspects of the instruments and monitor their state. These {\em auxiliary channels} include data from a variety of instrumental and environmental sensors, such as photodetectors and seismometers. These sensors can witness noise sources which couple to the interferometers. Their data can be used to diagnose and mitigate non-astrophysical couplings. An example is shown in Fig.~\ref{fig:channels}. 

A full, manual analysis of LIGO and Virgo auxiliary channel data is generally impracticable because of the huge number of instrumental and environmental monitoring sensors, amounting to several tens of thousand per interferometer. The power of ML to handle huge data sets proves invaluable in analyzing auxiliary channel data.

Methods for glitch identification based on auxiliary channel data have been extensively investigated within the LIGO and Virgo collaborations  \cite{2013PhRvD..88f2003B, 2013CQGra..30o5010E, 2011CQGra..28w5005S, 2010JPhCS.243a2005I,Colgan:2019lyo}. In these approaches, the GW channel is generally used to determine labels for the training samples while the glitch identification process relies only on information from auxiliary channels. Once the model is trained, it is fully independent of the GW channel data and considers only parameters computed from auxiliary channels known not to be related to astrophysical signals. 

The first study of canonical ML algorithms within a glitch detection framework such as Random Forests (RF), neural networks, and support vector machines, was published in Ref.~\cite{2013PhRvD..88f2003B}. Since then, multiple authors have investigated ML algorithms to infer the presence of non-Gaussian noise in GW data through features in auxiliary channels. 

iDQ is a glitch detection pipeline that can produce real-time data products in low-latency \cite{essick_thesis, Essick2020}. iDQ decomposes the problem of glitch identification into a 2-class classification scheme within a supervised learning framework with several asynchronous tasks: training, cross-validation, calibration, and low-latency prediction. The pipeline utilizes glitch features in auxiliary channels generated in real-time \cite{2004CQGra..21S1809C, Godwin2020} to construct supervised-learning training sets from recent data labeled by witnessing noise in the target channel, typically taken to be $h(t)$. It then automatically trains a variety of ML algorithms to identify glitches in the target channel. 
 
iDQ operated in real-time throughout the first three LIGO-Virgo observing runs, providing probabilistic statements about the presence of glitches in LIGO data and their auxiliary witnesses. 
 
iDQ's {\em Ordered Veto List} \cite{2013CQGra..30o5010E} algorithm contributed to the rapid release of the GW170817 BNS event by autonomously identifying the glitch coincident with the GW trigger in the LIGO Livingston detector \cite{PhysRevLett.119.161101} and in multiple auxiliary witnesses within 8 seconds of the event first being reported.

The LIGO Scientific Collaboration developed several additional promising approaches for auxiliary channel-based glitch identification in high-latency settings. These approaches typically use different auxiliary features, ML algorithms, and channel-sets. 

Two fast algorithms which aim at identifying the origin of glitches and can be used with minimal tuning to track their causes are based on RF and Genetic Programming (GP) \cite{Cavaglia:2018xjq}. RF and GP methods are interpretable, easy to use and tune, and can work with relatively small data sets without the inherent risk of overfitting. The algorithms require as  input a list of times when a specific class of noise transients occurs and rely on features which are directly drawn from the numerical metadata generated by real-time LIGO-Virgo data quality pipelines, such as Omicron \cite{Omicron}. This approach minimizes the feature generation step of the process, which is the typical bottleneck for ML-based glitch investigations. The data sets are assembled with features from noise transients and randomly-selected background triggers. The algorithms can be quickly trained and run on LIGO-Virgo computing clusters. A typical analysis with a number of noise transients of the order of a few thousands can be completed in minutes. The methods were validated in Ref.~\cite{Cavaglia:2018xjq} on two sets of $h(t)$ glitches with known origin from the first two LIGO-Virgo observing runs. 

Another ML tool that utilises auxiliary channel information is Elastic-net based ML for Understanding (EMU) \cite{Colgan:2019lyo}. EMU uses data from the {\em full} list of LIGO's auxiliary channels per detector site. The algorithm uses logistic regression with elastic net regularization \cite{Bishop,Hastie} to predict the probability of a glitch. Instances where a glitch is predicted to be present are classified as 1 (``glitch'') while instances where a glitch is predicted to be absent are classified as 0 (``glitch-free'') with a continuum of certainty between these limiting values. As other algorithms described above, EMU provides a measure of the auxiliary channel significance in predicting $h(t)$ glitches. This may enable researchers to uncover instrumental and environmental noise couplings to the GW data stream that can be used by commissioners to eliminate the instrumental root of the glitches. EMU's initial performance is illustrated in Ref.~\cite{Colgan:2019lyo}. It was also characterized using automatically clustered subsets of glitches, which can be derived according to frequency, duration, and other trigger generator parameters, or by using existing methods to identify classes of glitches \cite{2017CQGra..34f4003Z}. 

Ref.\ \cite{Mukund:2018occ} uses ML regression and clustering methods to infer peak ground velocities from Earthquake Early Warning alerts. The algorithm is trained on archival seismic data to determine the ground motion and the state of a GW interferometer during an earthquake. The estimated ground velocity is then used to forecast the potential effect of earthquakes on the detector in near real-time making it possible to switch the detector control configuration during periods of excessive ground motion.

{\em Hey LIGO} \cite{Mukund:2017fal} is an ML-based information retrieval tool which aims at supporting the commissioning and characterization efforts of GW observatories. The algorithm responds to an user query by searching the detector open source logbook data and returning information on detector operation, maintenance, and characterization tasks. The {\em Hey LIGO} web application incorporates a natural language processing-based information retrieval system that can also perform visualization of the user-queried data. 

\subsection{Methods for non-stationary noise subtraction and denoising}
\label{subsec:non-stationary}

ML techniques can be used to identify, model and subtract technical noises that couple in non-stationary or nonlinear ways. In particular, ML algorithms can be designed to construct filters for nonlinear noise subtraction. The output signal measured by LIGO and Virgo interferometers consists of the sum of three components (1) noise which can be removed, (2) noise which cannot be removed, and (3) GW signals. Unfortunately, the noise sources which can potentially be filtered out do not necessarily couple linearly into the interferometer. Therefore, even after Wiener filtering \cite{wienerbook}, there exists a substantial amount of nonlinear couplings polluting the output. Leveraging the ability of ML to infer nonlinear functions, environmental and control data streams can be used as input of neural networks to find the transfer functions of the systems producing nonlinear noise in the detector output. The trained network can then be used to subtract those nonlinear couplings from the output data and lower the total noise floor. 
 The above method was implemented, for example, in DeepClean \cite{DeepClean} and NonSENS \cite{VaHu2019}.

An especially interesting case is when the noise source is monitored by one or more available {\em fast} signals and its coupling can be described as linear on short time scales, but with coupling transfer functions that vary on longer time scales. The method can be illustrated by considering the effect of the longitudinal control noise due to the signal recycling cavity feedback, which couples to the detector strain in a frequency region that spans between about 10 and 300 Hz. The coupling is modulated on time scales longer than about 1 second by the residual angular motion of the mirrors. In this case and similar circumstances, it is possible to efficiently track the time-varying noise coupling using the interferometer angular control signals and develop a stable, parametric model of the noise polluting the detector strain \cite{VaHu2019}. This model can be used to perform a time-domain subtraction of the noise that outperforms the performance of any linear and stationary scheme. This non-stationary noise subtraction scheme was successfully applied to LIGO data during the first half of the third observing run. The method allowed removal of the non-stationary power supply line coupling that was limiting the detector sensitivity at the mains frequency of 60 Hz and in a 4 Hz-wide band around 60\ Hz due to sidebands created by the coupling modulation \cite{VaHu2019}.

Deep learning may also be used to uncover underlying signals by applying various denoising algorithms. Algorithms to denoise GW data include the total-variation method \cite{2014PhRvD..90h4029T,2018PhRvD..98h4013T} which applies the split Bergman regularization to obtain the total-variation regularization \cite{2015ASSP...40..289T}, dictionary learning \cite{2016PhRvD..94l4040T,alej2020application}, deep learning \cite{2020PhLB..80035081W} with  {\tt WaveNet} implementation \cite{2016arXiv160903499V}, and deep recurrent neural networks \cite{10.5555/553011,2013arXiv1312.6026P} in denoising auto-encoders architecture \cite{2018APS..APRS14008S,2019arXiv190303105S}. 

\section{Gravitational waveform modeling}
\label{sec:wf_model}

\begin{figure}[t]
    \centering
    \includegraphics[width=7.5cm]{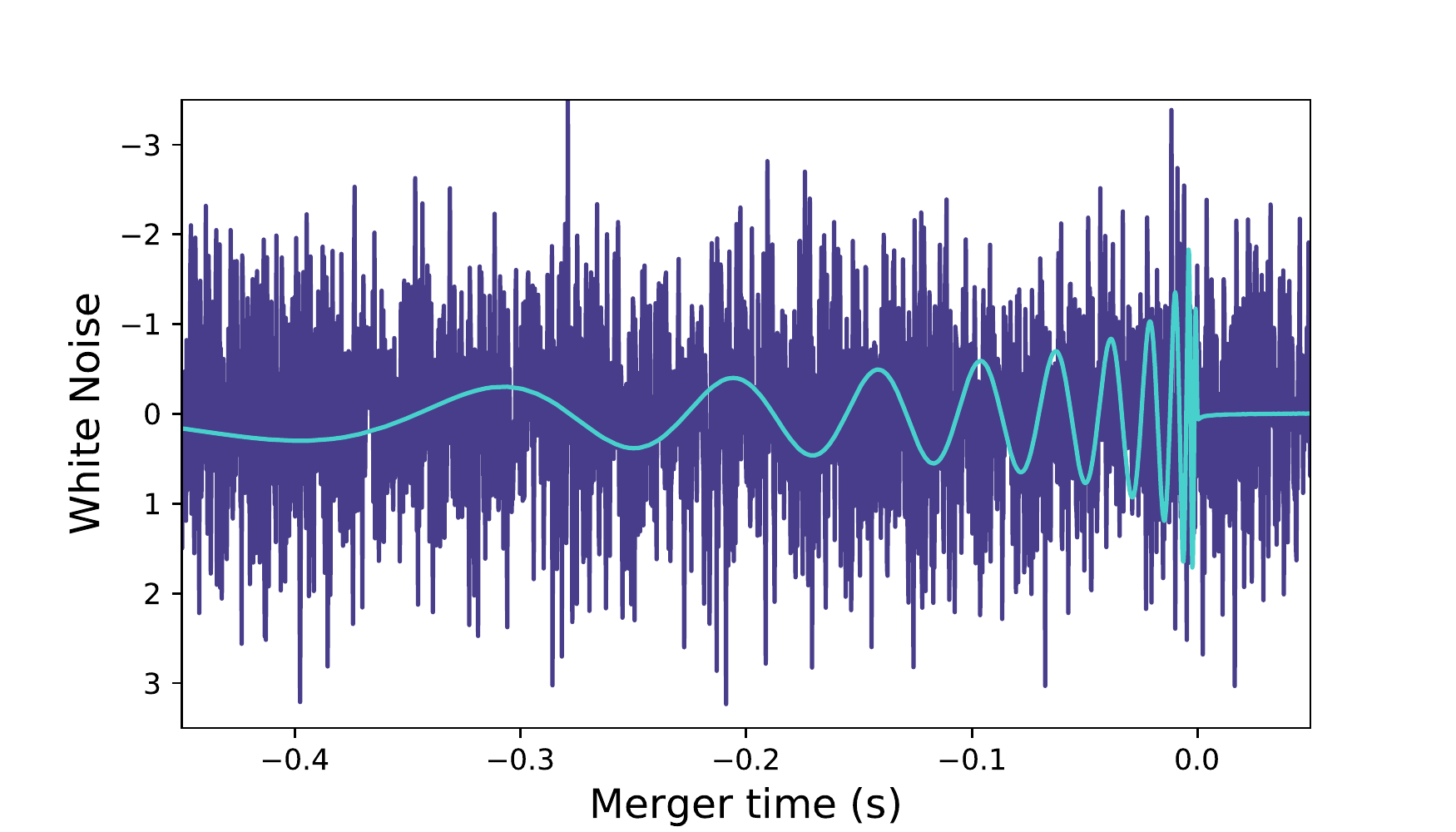}
    \includegraphics[width=7.5cm]{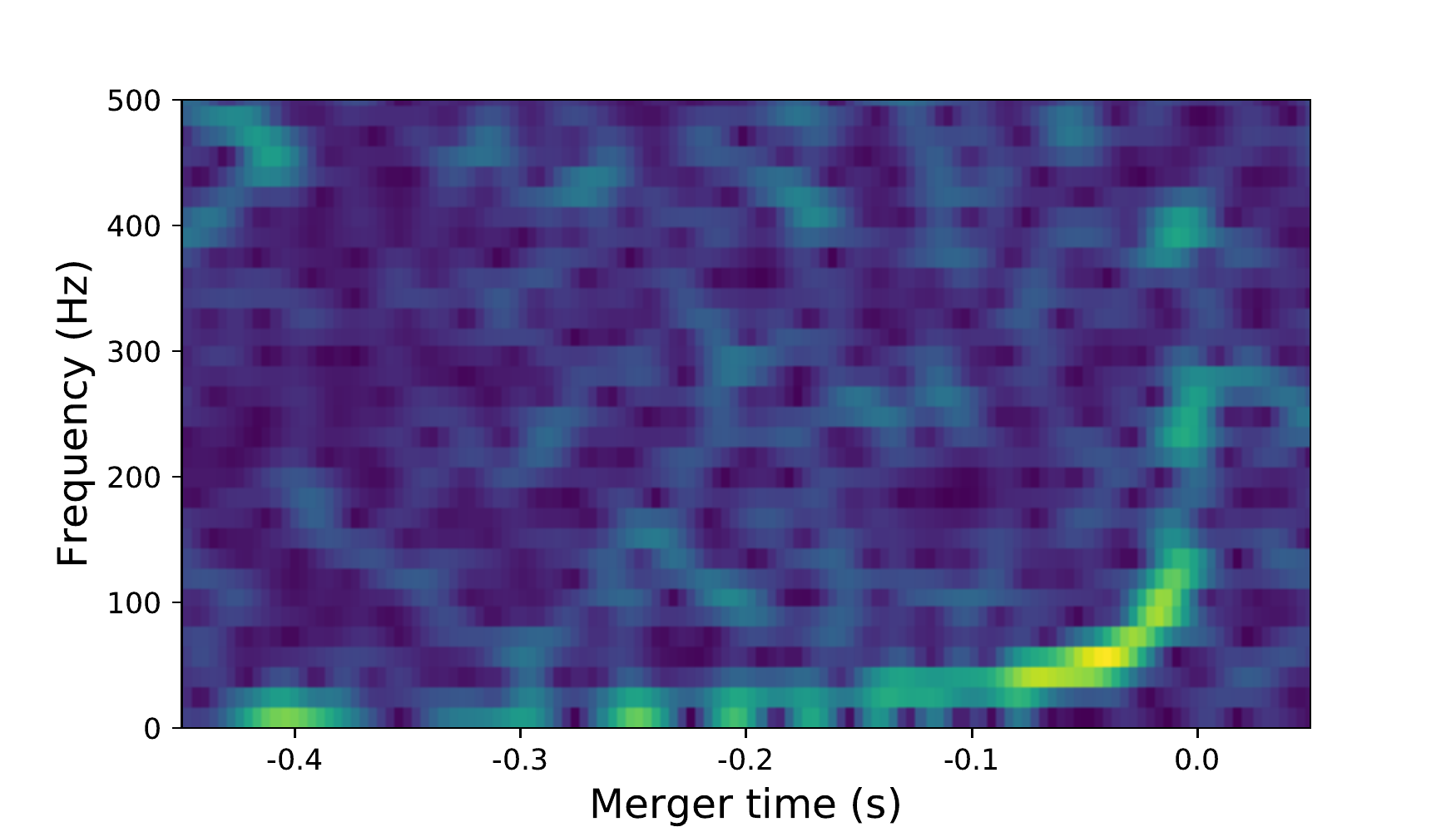}
    \caption{A typical BBH GW signal in whitened design sensitivity detector noise. The signal was generated with two 20 solar mass black holes and an optimal SNR=17. Left: Whitened time series. Right: Time-frequency spectrogram of the data showing the signal chirplike shape.}
    \label{fig:waveforms}
\end{figure}

LIGO and Virgo searches for GWs for CBC systems which rely on a matched filtering analysis~\cite{Nitz:2018rgo, Sachdev:2019vvd, Messick:2016aqy, Adams_2016, SPIIR} and the estimation of source parameters,performed with Bayesian inference~\cite{veitch:15,2019ApJS..241...27A}, require gravitational waveform templates. Waveform template are needed to compute the Signal-to-Noise Ratio (SNR) and the significance of GW triggers, and the posterior probability distribution of the signal parameters. Figure \ref{fig:waveforms} shows an example of a BBH signal in whitened detector data.

Accurate solutions of the Einstein equations for the two body problem can be obtained with Numerical Relativity (NR) simulations. However, high computational cost required to produce NR waveforms prevents the production of waveform catalogs spanning the full CBC parameter space~\cite{Pretorius:2005gq,Campanelli:2005dd,Baker:2005vv,Bruegmann:2006at,Centrella:2010mx,Mroue:2013xna,Jani:2016wkt,Healy:2017psd,Boyle:2019kee}.
The main parameters that describe a quasi-circular binary system of rotating BHs are the mass ratio and the spin vectors of the two objects. Neutron stars have additional internal degrees of freedom that are described by their tidal deformability, and generic compact binaries can also inspiral on eccentric orbits which further increases the dimensionality of the binary parameter space. 
Since binary parameters of GW events are a priori unknown their matched filter analysis with GW detector networks requires models of the emitted GWs that smoothly vary as a function of binary parameters, rather than solutions at isolated points as provided by NR simulations.

It is important to emphasize that waveform models need to be accurate and computationally efficient to use them in detection and parameter estimation pipelines. Good accuracy in terms of the overlap, the normalized cross-correlation maximized over time and phase shifts, between a waveform model and the most accurate waveforms available (usually NR simulations, or NR simulations stitched together with inspiral waveforms) is crucial to avoid missing events in searches or mis-measuring binary parameters. On the other hand, waveform models need to be fast to evaluate since searches and parameter estimation require tens to hundreds of millions of waveform evaluations and the cost rises for lower mass binaries with hundreds of GW cycles in band.

The coalescence of a binary system starts with an inspiral phase corresponding to a large separation of the two objects where the GWs emitted can be computed with the post-Newtonian (pN) formalism~\cite{Blanchet:2013haa}.
As the objects come closer, the orbital velocity becomes comparable with the speed of light and the pN expansion breaks down. In this strong gravity regime NR simulations are needed to compute the emitted gravitational radiation during the merging of the objects.
If the remnant is a BH the final object relaxes to a Kerr BH under the emission of a sum of damped quasi-normal modes~\cite{Berti:2009kk} in the so-called ``ringdown'' phase.
 
LIGO and Virgo rely on approximate solutions that are traditionally obtained through the effective-one-body (EOB) or phenomenological modeling approaches.
In the EOB formalism~\cite{Buonanno:1998gg,Buonanno:2000ef} the PN inspiral information is re-summed and calibrated to NR data, and the merger-ringdown part is obtained from a phenomenological fit to NR data. EOB models~\cite{Buonanno:1998gg,Buonanno:2000ef,Damour:2000we,Damour:2001tu,Damour:2008gu,Pan:2011gk,Taracchini:2012ig,Taracchini:2013rva,Pan:2013rra,Damour:2014sva,Nagar:2015xqa,Bohe:2016gbl,Babak:2016tgq,Knowles:2018hqq,Cotesta:2018fcv,Nagar:2018zoe} first solve the orbital dynamics by solving a complex system of ordinary differential equations and then obtain the waveform as a second step.
Phenomenological models~\cite{Ajith:2011,Santamaria:2010yb,Hannam:2013oca,PhenomPv2_tech_rep,Husa:2015iqa,Khan:2015jqa,Mehta:2017jpq,London:2017bcn,Khan:2018fmp,Khan:2019kot,Pratten:2020fqn} provide full gravitational waveforms by tuning an extended PN inspiral expansion and separate analytical functions for the merger and ringdown waveforms to NR simulations that have been combined with PN or EOB inspiral waveforms, and then smoothly combining these three regions. 
Surrogate or reduced order models of NR or EOB waveforms~\cite{Field:2013cfa,Purrer:2014fza,Purrer:2015tud,Blackman:2015pia,Bohe:2016gbl,Blackman:2017pcm,Blackman:2017dfb,Lackey:2018zvw,Doctor:2017csx,Varma:2018mmi,Varma:2019csw,Williams:2019vub} can significantly accelerate these waveforms while keeping high accuracy. They have become indispensable tools for GW data analysis over the past several years. Surrogate models fit interpolated decomposed waveform data pieces over the binary parameter space and also compress the waveform in time or frequency.

The above modeling approaches rely mostly on traditional polynomial interpolation or fitting techniques for model construction. In the past few years, more advanced ML techniques have started to be explored for model building.
Gaussian process regression (GPR) can be considered as a generalization of a multivariate normal distribution and allows to not only build a model of input data, but at the same time obtain a measure of the uncertainty in the data. GPR enables us to infer the waveform at points of the parameter space not covered by numerical relativity by assuming a joint Gaussian distribution between the known and predicted values, then computing the posterior probability of the predicted parameters. 
GPR fits have been used to build surrogate models of non-precessing BBH systems with iterative refinement~\cite{Doctor:2017csx}, as well as surrogates of precessing binary NR waveforms~\cite{Varma:2019csw, williams2019precessing}. 
The usefulness of including the uncertainty on the waveform in models is the capability to marginalize over waveform uncertainty in Bayesian estimations of the source parameters~\cite{Moore:2014pda,Moore:2015sza}.\par
A comparative study of regression methods used for fitting or interpolating waveform data in waveform models covering both traditional polynomial methods, GPRs, and, for the first time, artificial neural networks has been carried out recently~\cite{Setyawati:2019xzw}.
The increasing use of ML in GW source modeling motivated a study
comparing the performance of ML approaches against traditional
regression methods. In \cite{Setyawati_2020}, the authors carefully investigated the accuracy, training, and execution time of ML methods (GPR and ANN) against linear interpolation, radial basis functions, tensor product interpolation, polynomial fits, and greedy multivariate polynomial fit. This study
addressed the question of whether a more sophisticated and complicated method is necessary to build a gravitational waveform model for aligned-spin and precessing binaries. They concluded that sophisticated regression methods, especially ML, are not necessarily needed in standard GW modeling applications, although ML techniques might be more suitable for problems with higher complexity, like fully spinning black holes.

A key input in the design of waveform models is the mass and spin of the remnant BHs, entirely predicted from the initial BHs parameters by general relativity and computed in numerical simulations.
The parameters of the remnants from non-precessing binary systems are traditionally determined with explicit fits to numerical relativity results~\cite{Jimenez-Forteza:2016oae, Healy:2016lce}, a procedure that has been extended to determine the final spin magnitude for precessing systems~\cite{Hofmann:2016yih}.
The description of fully spinning BHs however requires a larger dimensional space, where ML methods are particularly suited to capture the complex relationship between the initial and final parameters. For this reason, the determination of remnant BHs mass, spin and recoil velocity for precessing systems has been performed independently with GPR and Deep Neural Networks (DNN), both methods showing increased accuracy compared to existing fits~\cite{Varma:2018aht,Haegel:2019uop}. 

GWs from the remnants of BNS mergers can be used to place constraints on the neutron star equation of state. However, the NR simulations that are currently used to model the merger and post-merger stage of such GW signals are computationally expensive to generate and of limited accuracy, thereby restricting the ability to use them to perform parameter estimation of candidate GW detections. A hierarchical ML algorithm trained on NR simulations was developed in Ref.~\cite{Easter2019} that can quickly generate gravitational waveform amplitude spectra with mean overlaps of $\gtrsim\,0.95$, and can also be used to place constraints on the quadrupolar tidal deformability given sufficient SNR in the post-merger signal.

\section{Gravitational-wave signal searches}
\label{sec:searches}

Searches for GW signals in ground-based detector data are typically split into four different types, depending on their search strategy. The first type includes CBCs. BBH, BNS and NSBH are the most common type of sources and the only type of source detected to date. The second type, which we refer to as GW ``bursts'', includes transient sources with an unknown or partially modeled signal morphology, as for example, the ones produced by CCSNe. The third source type includes long-duration, continuous GWs that may be produced by an individual rotating neutron star. The fourth type is a stochastic background of GWs, which could consist of remnant GWs from the Big Bang or distant unresolved CBCs. In this section, we review ML approaches that have been developed to enhance LIGO and Virgo searches for these different types of GW signals.
\subsection{CBC searches}
\label{subsec:cbc}
Matched filtering \cite{wienerbook} is a common approach to search for sources from CBCs in LIGO and Virgo data \cite{Abbott_2020}. Matched filtering works by first taking as input the raw calibrated strain data. Then a {\it{template bank}} of waveforms is generated spanning a large astrophysical parameter space. 
The matched filter searches then produce a list of GW triggers by cross-correlating the GW data with the waveforms in the template bank divided by the interferometer's spectral noise density.
Signal consistency tests are done on triggers above a certain SNR threshold generated from the matched filtering algorithm in order to determine the time-frequency distribution of power in a trigger. This distribution is compared to what would nominally be expected from the power in the matched filtering waveform. The comparison is done by splitting the template up into a number of frequency bins, such that each bin contributes an equal amount of power to the total matched filter SNR. An additional statistic is then constructed to compare the expected to the measured power in each bin \cite{PhysRevD.71.062001}.

Recently, there have been developments in exploring how random forest ML algorithms may be used as an alternative to standard CBC signal detection techniques. 
Using a ranking statistic derived from a random forest of bagged decision trees it was reported that sensitivity improvements were achieved 
on the order of $70_{\pm13}\%-109_{\pm12}\%$ compared to matched filtering \cite{2015PhRvD..91f2004B}. This study was carried out as an intermediate mass BH and a stellar-mass BBH search $(\gtrsim 25\,M_{\odot})$. In Ref.~\cite{2017PhRvD..96j4015K}, the random forest classifier was instead trained/tested on simulated single-detector NSBH events and the authors use hand-crafted features derived from a bank of inspiral templates as input. The classifier was able to detect $1.5-2$ times as many signals as those those found by standard matched filter detection techniques at low false positive rates as compared to the standard “reweighted SNR” statistic, and does not require the chi-squared test to be computed. The results from both of the studies discussed above, show that random forest ML approaches have the potential to produce higher detection efficiencies for CBC GW sources.

Other promising approaches to CBC signal identification have been proposed by several groups by applying deep learning-based methods \cite{Kim_2015, Kim_2020, PhysRevLett.120.141103, 2018PhLB..778...64G}. CNN algorithms have been applied in multiple studies towards the search for GWs from CBC signals. CNNs were applied to simulated BBH signals and real LIGO events in non-Gaussian noise, as well as non-Gaussian noise alone in an attempt to classify such signals \cite{George:2017pmj}. This analysis reported that CNNs perform as well as the standard matched filter method in extracting BBH signals under non-ideal conditions. CNNs and matched filtering 
were also compared in Gaussian noise, where matched filtering is thought to be ideal \cite{PhysRevLett.120.141103} and were able to match the efficiency of a matched filter analysis. These studies address the fundamental question of the feasibility of deep learning application to CBC GW searches \cite{PhysRevLett.120.141103, 2018PhLB..778...64G}.

A different conclusion is reached in Ref.~\cite{Gebhard:2019ldz}, where CNN methods for CBC searches are shown not to provide an accurate measure of the detection statistical significance. The latter can be achieved in matched filter searches by sliding the data of one detector in time by an amount which is larger than the typical GW travel time between detectors. A measure of coincident events after the time slides is used to estimate the search background and the significance of detection candidates. CNN algorithms should implement an accurate statistical measure of the background to take current, ML-based CBC searches from proof-of-principle studies to production search codes.

\subsection{Burst searches}
\label{subsec:burst}

A GW burst is a short duration signal with an unknown or partially modeled waveform morphology due to complicated or unknown astrophysics. Potential sources of GW bursts are CCSNe \cite{Fryer2003}, pulsar glitches \cite{PhysRevLett.87.241101}, NSs collapsing into BHs \cite{Baiotti_2007}, cosmic string cusps \cite{PhysRevD.71.063510}, and many others. The uncertainties in signal morphology make it difficult to produce training sets for GW burst signals. 

One of the standard LIGO-Virgo GW burst search methods is the coherent Wave Burst (cWB) pipeline \cite{2016PhRvD..93d2004K, 0264-9381-25-11-114029}. The cWB algorithm uses a Wilson-Daubechies-Meyer wavelet transform to measure the excess power in the time-frequency domain that occurs coherently between multiple detectors. A maximum likelihood approach is then applied to determine the probability of a signal being present in the data and produce a list of GW detection candidate triggers. This model-independent method is well-suited for searching for signals with unknown morphology. The sensitivities of burst searches are generally more affected by short-duration glitches that may occur coincident in time between multiple detectors. No detections other than CBC sources have been reported by cWB unmodeled searches to date  \cite{2017PhRvD..95d2003A}. 

Currently there are no published ML searches for generic GW bursts. However, there have been ML searches for specific burst source types. In Ref.~\cite{2017CQGra..34i4003V}, the authors employ a neural network algorithm to reduce the impact of glitches on the cWB burst search and increase the significance of the CBC signals which are detected by the pipeline.

In recent years, multi dimensional simulations of CCSNe have produced a selection of GW signal predictions from CCSN explosions \cite{2019MNRAS.487.1178P, 2019ApJ...876L...9R, 2019MNRAS.486.2238A, 2018MNRAS.475L..91T}. However, some of these simulations include approximations of the required input physics that may result in artificial changes in the GW emission. Moreover, many simulations are ended before the peak GW emission time due to lack of computational resources. Despite these issues, common features in the time-frequency GW emission have recently been identified by various CCSN simulation codes. This has allowed researchers to produce approximate models for a wider range of the CCSN parameter space than can be explored with full CCSNe multi-dimensional simulations \cite{2019MNRAS.482.3967T}. These approximate models allow us to explore supervised ML techniques for CCSN searches. 

In Ref.~\cite{2018PhRvD..98l2002A}, the authors apply a CNN to searches for GW bursts from CCSNe. Training and testing are performed with a parametrized phenomenological waveform model which is designed to match the most common features predicted by CCSN simulations. The method uses 100 different parameterizations of the phenomenological model. cWB pre-processing data is used to prepare images which are fed into the CNN. Red-green-blue (RGB) images are produced to determine the number of detectors where a signal is present: Red (R) for LIGO Hanford, green (G) for LIGO Livingston and blue (B) for Virgo. The algorithm is shown to improve the efficiency of cWB in its standard configuration.

A technique to reduce the background of searches for galactic CCSNe in single-interferometer configurations was developed in Ref.~\cite{Cavaglia:2020qzp}. The method is based on a supervised evolutionary algorithm, Genetic Programming (GP). The procedure assumes that the event time and the distance of the CCSN are known from neutrino and optical observations. The GP algorithm is first trained on off-source data to produce a multivariate expression of the trigger features which is used as a cut to lower the search background. The multivariate expression can then be applied to on-source windows around a GW event candidate to increase the detection confidence. The effectiveness of the method was tested  by injecting the set of waveforms used in the latest LIGO-Virgo CCSN search into 1.47 days of O1 data. The features are extracted from the standard cWB pipeline. The GP algorithm is then used to classify the triggers and remove the noise triggers. The outcome of the procedure is an increased statistical significance of GW candidate triggers that leads to a reduction of the SNR needed for a detection at 3$\sigma$ confidence level by a factor $\sim$ 3.

In Ref.~\cite{10.1088/2632-2153/ab7d31}, the authors train a CNN on waveforms obtained by 3D simulations of neutrino-driven CCSNe, phenomenological sine-Gaussian waveforms, and scattered light glitches. A wavelet detection filter (WDF) \cite{8553393} with time-domain whitening \cite{Cuoco_2001} is used to extract GPS triggers from the simulated Virgo and Einstein Telescope noise backgrounds. Whitened time series and spectrograms are used as input of 1D and 2D CNN algorithms to classify signal and noise classes. The method is then tested on CCSN models removed from the training sets. In a multi-label classification scheme, the CNN is capable of distinguishing among the individual CCSN and glitch classes at different SNRs. In Ref.~\cite{2019arXiv191213517C}, the authors train a CNN algorithm on the time series of CCSN waveforms and classify signals with two different explosion mechanisms.    

\subsection{Continuous wave searches}

Narrow-band Continuous GWs (CWs) from spinning deformed neutron stars~\cite{Prix:2009oha}
have not yet been observed in data recorded by LIGO and Virgo.
Although the expected waveform signatures for this type of signal are well known, their small amplitude makes them extremely hard to detect, and any search needs to process long stretches of data.
As the number of parameter space points to search grows with the observation time, the sensitivity of current searches is limited by the available computing power.
Several established search methods exist with different trade-offs between sensitivity, robustness and computational efficiency.
For searches covering wide parameter spaces,
analysis pipelines usually follow a hierarchical approach~\cite{Brady:1998nj} with multiple stages.
The first stage evaluates a detection statistic over a dense grid of candidate waveforms.
Additional stages often include clustering of candidates to reduce computational waste, vetoing of known types of instrumental artifacts, and follow-up with increased resolution and/or observation time. See Ref.~\cite{Riles:2017evm} for a recent review.

The huge computational cost of conventional CW searches makes ML approaches a promising alternative as they are fast at searching once training has been completed.
In addition, many current follow-up methods also involve manual tuning steps, for which extensive simulation campaigns must be repeated on each new data set and search setup~\citep[e.g ][]{Ming:2017anf,Walsh:2019nmr}. Robust and flexible ML solutions would reduce this additional human and computational effort.

At this point, several teams have started to explore two avenues of using ML methods in CW searches:
(i) using ML algorithms as drop-in replacements for parts of a CW analysis pipeline that is still based on traditional grid search methods for the initial search stage;
(ii) full ML analysis of the initial strain data.

Among the first category, Ref.~\cite{Morawski2019DeepLC} presents an application of deep learning
in the classification of CW signal candidates.
A conventional initial search, based on the $\mathcal{F}$-statistic matched filter pipeline - {\tt TD-Fstat search} \cite{polgraw-allsky} (see the documentation in \cite{polgraw-allsky-docs}), generates a large number of multi-dimensional candidate signal distributions that have to be further analyzed.
Then 1D and 2D versions of a CNN classifier are implemented, trained, and tested on a broad range of signal frequencies, corresponding to the reference frequency of the narrow-banded data. 
The training set contains Gaussian noise, simulated CWs from spinning triaxial-ellipsoid neutron stars, and stationary lines mimicking detector artifacts~\cite{Covas:2018oik}.
The authors of \cite{Morawski2019DeepLC} demonstrate that these CNNs correctly classify the instances of data at various SNRs and frequencies, while also showing concept generalization, i.e., satisfactory performance at frequencies the networks were not trained on.

For another $\mathcal{F}$-statistic search pipeline,
the hierarchical semi-coherent analysis~\cite{Pletsch:2008gc,Pletsch:2009uu,Pletsch:2010xb}
running on the distributed computing Einstein@Home project~\cite{EatH},
Ref.~\cite{Beheshtipour:2020zhb} employs a deep learning replacement
for traditional clustering methods~\cite{Behnke:2014tma,Papa:2016cwb,Singh:2017kss}.
Clustering has the purpose of reducing the number of candidates from the initial search stage, enabling deep follow-up~\cite{Papa:2016cwb,Walsh:2019nmr} at acceptable computational cost, thus improving the sensitivity of the overall pipeline.
In \cite{Beheshtipour:2020zhb}, they train a region-based CNN~\citep[R-CNN, ][]{2013arXiv1311.2524G,2017arXiv170306870H}
on real output of an Einstein@Home search~\cite{Abbott:2017pqa} on Advanced LIGO O1 data.
They demonstrate a high detection efficiency at low false
alarm rate for sufficiently strong signals, and investigate the scaling of this performance with signal strength.
They identify the R-CNN's brittle response to instrumental disturbances in the data as both a problem and an opportunity;
since in the current implementation it already distinguishes these from normal noise, they expect that a more completed classification into three classes of normal noise, disturbances and signals
can be pursued.

In a more radical approach than the works above, the authors of Ref.~\cite{Dreissigacker:2019edy} apply a deep neural network to search for CWs from unknown spinning neutron stars on the raw time-series data.
A CNN is trained on Gaussian noise with signal injections
and compared to a matched filter search~\cite{Wette:2018bhc}.
The analysis shows that the method is competitive with matched filtering, at least under these idealized noise conditions
and when using data spans of limited duration.
Thus, they provide the first demonstration of a full-ML search for CWs without prior input from a traditional first-stage search.
However, the authors themselves consider this as only a `first proof-of-principle' and point out several steps required for developing it into a mature analysis pipeline, including the use of data from multiple GW detectors and dealing with non-Gaussian real detector data, including the pervasive narrow spectral artifacts~\cite{Covas:2018oik} that have long been a challenge to traditional CW search methods \cite{Keitel:2013wga,Leaci:2015iuc}.

Besides CNNs and other neutral networks, several other non-traditional approaches to CW searches have been recently explored.
One successful application is the hidden Markov model tracking CW search method~\cite{Suvorova:2016rdc,Suvorova:2017dpm,Sun:2017zge,Sun:2018owi}, where the true intrinsic emission frequency of a GW source is treated as a hidden variable tracked by signal frequencies in the detector data, thus allowing for deviations from a simple signal model or even for intrinsic frequency fluctuations (e.g., spin wandering in binary sources and pulsar timing noise).
The most likely time-frequency tracks are found with the Viterbi algorithm~\cite{Viterbi1967}. This method is different from most algorithms usually considered as ML, for example in that it does not require a separate training stage, but is nevertheless inspired by work in the computer and information science fields and provides another example for fruitful imports into GW science.
An independent Viterbi-based search has also been developed in Ref.~\cite{Bayley:2019bcb}

Methods originally developed for CW searches have also proved to be useful for long-duration transient GWs~\cite{Abbott:2018hgk,Keitel:2019zhb}.
After the detection of GW170817, several CW search methods have been adapted to search for transient GW signals from long-lived BNS merger remnants~\cite{Abbott:2018hgk,Miller:2018rbg,Sun:2018owi,Oliver:2019ksl}.
In this scenario, as for the shorter-duration post-merger signals briefly discussed above in Sec.~\ref{sec:wf_model}, waveforms are not well known, and hence these traditional template-based search methods are limited in robustness to deviations from their assumed models.
The authors of Ref.~\cite{Miller:2019jtp} have developed a CNN analysis for BNS post-merger signals lasting $\mathcal{O}($hours-days). They characterized the CNN approach, answering questions like:
(i) how to train them, (ii) how many samples to train on, (iii) robustness to signals on which the networks were not trained, and (iv) the effect of different architectures on detection efficiency and false alarm probability.

They also applied their CNN in a real search on the week of LIGO data after GW170817, producing upper limits on possible GW emission similar to those in~\cite{Abbott:2018hgk}.

Other research areas where conventional CW methods reach their limits in terms of (i) robustness of the underlying signal model, (ii) computational efficiency, (iii) adaptability to new situations with reasonable human time investment, include very young sources with rapid frequency evolution (similar to the BNS post-merger case),
glitching pulsars~\cite{Keitel:2019zhb}, and GW emission from neutron stars in binaries~\citep[e.g.][]{Meadors:2015vpc,Abbott:2019uwg,Covas:2020nwy}. Deep learning and other ML methods, or simple yet innovative methods imported from the computational science community,
like the Viterbi algorithm~\cite{Viterbi1967,Suvorova:2016rdc,Suvorova:2017dpm,Sun:2017zge,Sun:2018owi,Bayley:2019bcb}, can lead to big steps towards first detections of these elusive GW signal types.

\subsection{Stochastic background}

A stochastic GW background may consist of many CBC sources that are too distant to be individually resolved or remnant GWs from the Big Bang \cite{2019RPPh...82a6903C}. Typical searches for a Gaussian stochastic background apply the cross-correlation method described in Ref.~\cite{PhysRevD.59.102001}. For non-Gaussian stochastic backgrounds from stellar mass BBHs, a Bayesian nested sampling method can be implemented for optimal search sensitivity \cite{2018PhRvX...8b1019S}. In Sect.~\ref{sec:pe} we describe how ML can be used to improve the speed and accuracy of this method. A Gaussian mixture model is used to predict the discovery time of the GW stochastic background in Ref.~\cite{2018PhRvX...8b1019S}.

\section{Astrophysical interpretation of gravitational-wave sources}
\label{sec:pe}

\subsection{Parameter estimation}

To understand the astrophysics behind sources of GWs, it is essential to accurately measure the parameters of the source. For CBC signals, this is currently achieved using the \texttt{LALInference} \cite{veitch:15}, \texttt{Bilby} \cite{2019ApJS..241...27A}, or \texttt{RIFT} \cite{gwastro-PENR-RIFT} tools designed for Bayesian parameter estimation and model selection. 
A Bayesian framework allows us to calculate posterior probability density functions for the parameters of GW signals. It also allows us to calculate the evidence for different models which can be used for model selection. Bayesian evidence is computationally costly. In the case of CBC signals, this is due to the high number of signal parameters $(\sim 15)$, the process of generating waveforms, and the SNR and length of the GW data being analyzed. In \texttt{LALInference}, the computational issues are addressed using either nested sampling \cite{2010PhRvD..81f2003V} or Markov Chain Monte Carlo (MCMC) techniques \cite{Blasco2017}. 

To estimate a posterior distribution, MCMC techniques work by stochastically wandering though a parameter space, distributing samples that are proportional to the density of the posterior.  The \texttt{LALInference} implementation of MCMC uses the Metropolis-Hastings algorithm that requires a proposal density function to generate a new sample that can only depend on the current sample. The efficiency depends on the choice of the proposal density function.

Nested sampling is used to calculate the Bayesian evidence and can also produce posterior distributions for the signal parameters. Nested sampling transforms the multi-dimensional evidence integral into a one dimensional integral over the prior volume. First a set of live points are distributed over the entire prior. The point with the lowest likelihood is then removed and replaced with a point with a higher likelihood and continues until some stopping condition is reached. Posterior samples can then be produced by re-sampling the chain of removed points and current live points according to their posterior probabilities. This method can take days to weeks to measure the parameters of a GW signal. Speeding up this process becomes more important as GW detectors become more sensitive and more signals need to be analysed. 

Some efforts at speeding up GW parameter estimation with ML have already been implemented in LIGO and Virgo data analysis codes. RIFT is a generic algorithm that uses Gaussian process or random forest regression to approximate marginal likelihoods appearing in Bayesian inference for compact binary parameters. By efficiently parallelizing, RIFT can produce inferences at a much faster speed \cite{gwastro-PENR-RIFT-GPU}. Multinest \cite{2009MNRAS.398.1601F} is a generic algorithm that implements the nested sampling technique using ellipsoids. Bambi \cite{2012MNRAS.421..169G} incorporates the nested sampling in Multinest and combines it with a neural network to learn the likelihood function on the fly. It has been shown that Bambi can produce results comparable to the full nested sampling techniques at a much faster speed. 

Other efforts to perform parameter estimation directly with neural networks have been carried out in \cite{2019arXiv190301998S, 2018PhLB..778...64G,1909.06296,PhysRevLett.124.041102}. In Refs.~\cite{1909.06296,PhysRevLett.124.041102}, it was first shown that a neural network could reproduce GW Bayesian posteriors with both studies being carried out using independent methods. In Ref.~\cite{1909.06296}, a conditional variational autoencoder (CVAE) is used in order to approximate the Bayesian posterior given a GW time series. The neural network used in this work is derived by starting from the assumption that we would like to minimize the cross-entropy between the true Bayesian posterior and an approximate Bayesian posterior produced by the neural network. After some further derivations,  one arrives at a network configuration which, when given a GW time series after training, will reproduce the true $n$-dimensional Bayesian posterior in less than a few milliseconds. Five compact BBH parameters are inferred  ($m_1$,$m_2$,$d_l$,$t_0$,$\phi$) where phase ($\phi$) has been internally marginalized out and all remaining parameters are fixed. Other work has also been carried out in parallel by \cite{PhysRevLett.124.041102} using a multivariate Gaussian posterior model to produce 1- and 2- dimensional posteriors. This approach also utilizes reduced order modeling \cite{PhysRevLett.122.211101} in order to simplify the input GW time series to the network. Once trained on a generated set of waveforms with many different noise realizations, it is then easy to sample from the model with comparable latency to \cite{1909.06296} ($\sim 1-2 \: \mathrm{milliseconds}$).

In addition, Ref.~\cite{2020arXiv200207656G} uses a different method, known as normalizing flows, to produce GW posteriors  comparable to those in Ref.~\cite{2019arXiv190301998S,PhysRevLett.124.041102} in less than 2 seconds. 
A normalizing flow is a series of invertible transformations that can be used to transform a simple initial distribution (in this case, a multivariate Gaussian) into a more complex target distribution. Normalizing flows are usually realized as neural networks, meaning that the parameters of the transformations are learned. The authors then compare three different approaches: a conditional variational autoencoder (similar to the network used in \cite{2019arXiv190301998S}), a masked autoregressive flow (MAF) \cite{1705.07057}, and a  CVAE with autoregressive flows. The results from both the single MAF and the CVAE networks are comparable, while the combined CVAE and MAF model produce the optimal result.

All of the current GW ML parameter estimation studies are still at the proof-of-principle stage, but they show that ML will be a promising tool for future GW parameter estimation. As the detectors sensitivity improves, the number of detections will significantly increase, and one of the advantages in using these ML techniques will be the speed in measuring the sources astrophysical parameters with respect to traditional methods, making it easier to process a large number of GW alerts.

\subsection{Low-latency source properties inference}
\label{sec:llinf}

The source property inference of CBC events is one of the three astrophysical data products that the LIGO and Virgo collaborations provide to the external community, the other two being 3D-sky localization \cite{Singer_2016} and source classification \cite{Kapadia:2019uut}. The source-property inference, also known as {\em EM-Bright}, consists of two statistical metrics: (1) the probability that the CBC system contains a neutron star of mass less than $3.0\,M_{\odot}$, \texttt{P(HasNS)}, and (2) the probability that the final coalesced object is surrounded by tidally-disrupted matter after the merger, \texttt{P(HasRemnant)}. Numerical solutions of the Einstein equations in the presence of matter provide us with estimates of the amount of tidally-disrupted matter. Several fitting formulae  have been derived to compute this quantity in low-latency \cite{Foucart:2012nc, Foucart:2018rjc}.

In an ideal situation, Bayesian parameter estimation of GW data can provide the posterior probability distribution of the various source parameters. Cuts based on the maximum neutron star mass can then be applied to compute \texttt{P(HasNS)}. Similarly, the fitting formula for the tidally-disrupted matter can be applied to the parameter posterior distributions to infer \texttt{P(HasRemnant)}. However, the fastest LIGO-Virgo parameter estimation infrastructures currently available do not meet the low-latency requirements for electromagnetic (EM) follow-up in X-ray and optical wavelengths. Getting the most reliable EM-Bright values in absence of Bayesian posterior samples poses the most important challenge in source properties inference.

Matched filter searches give point estimates of masses and spins that can be used for low-latency estimates of EM-bright properties. Individual mass and spin components from matched filter searches have often large errors with respect to true parameter values. During the O2 run, LIGO and Virgo researchers estimated these errors by implementing a Fisher approximation method to construct an ellipsoidal region of the parameter space around the matched filter point estimates. However, this method is only effective for high SNR signals and it ignores the matched filter search biases in parameter measurements. A supervised ML algorithm was developed to address this issue in Ref.~\cite{Chatterjee:2019avs}. The algorithm uses the scikit-learn \cite{scikit-learn} KNeighborsClassifier method to classify a source as HasNS or HasRemnant. An input data set is created by injecting simulated signals into GW data and performing a search. A map between true values and matched filter search point estimates is then used to train the classifier. Source property inference of GW detection candidates obtained with this method were provided in low-latency to the astronomy community during O3. 

\subsection{Rates and populations of gravitational-wave sources}

LIGO and Virgo have already detected a small population of CBC sources \cite{PhysRevX.9.031040}. The number of detections is expanding rapidly as the detectors become more sensitive, and within the next few years the population size will reach into the hundreds, allowing us to perform detailed CBC population studies. In the case of compact binaries, measuring properties such as their mass and spin distributions could allow us to determine their formation mechanism \cite{0264-9381-27-11-114007, 2017Natur.548..426F, 2017MNRAS.471.2801S, 2015ApJ...810...58S, 0264-9381-34-3-03LT01, 2017PhRvD..95l4046G, 2018arXiv180102699T}.

Several studies have demonstrated how ML can be applied for population analysis once a large enough population of CBC sources has been detected. A few of these studies apply unmodeled clustering techniques, such as Gaussian Mixture Models, to determine if GW detections come from multiple CBC populations \cite{2017MNRAS.465.3254M, 2017arXiv171202643W, 2019arXiv190504825P}. They perform the clustering on the mass and spin measurements of the individual CBC detections. This method is well-suited as it does not require any prior knowledge of the expected CBC populations. This allows us to determine how many different populations of sources are present in detections made by LIGO and Virgo, and what fraction of detected events belong to each population. The masses and spin distributions of each population can then inform us of the differences in formation of the different source populations. 

One of the current standard methods for population analysis in GWs is Bayesian hierarchical modeling. This approach is effective when there are trusted known population models, and allows for mixing ratios of different populations. However, this method can be computationally expensive. In \cite{2020arXiv200209491W}, the authors combine hierarchical Bayesian modeling with a flow-based deep generative network. Combining ML with Bayesian hierarchical modeling allows for a population analysis that is too computationally complex for hierarchical modeling alone. 

These modeled and unmodeled ML techniques will be applied to the GW populations detected in the future GW detector observing runs.

\subsection{Identification of electromagnetic counterparts to gravitational-wave sources}

Detecting EM emission from a GW source can considerably increase the knowledge of the astrophysical source properties and contribute to the validation of cosmological models. The authors in \cite{PhysRevD.100.103025} applied an Artificial Neural Network (ANN) to localize GW signals from BBH mergers. The input data to the ANN consists of the arrival time delays, amplitude ratios, and phase differences from multiple GW detectors. For the purpose of GW source localization, they divided the sky into grids or sectors. The samples were labelled with the sector number to which they belonged and the ANN was trained to classify them into their correct sector. It was found, when testing their method with simulated BBH signals in Gaussian Advanced LIGO noise, that for coarse angular resolution (18-128 sectors), the model is able to classify unseen GW samples into their correct sector in more than 90\% of cases, when a multi-labelling scheme is applied. For finer angular resolution (1024-8192 sectors), the exact classification accuracy drops. For these cases, the probability distribution of the sectors, obtained for each GW sample, was used as a ranking statistic to calculate the areas of 90\% probability contours. This method is potentially orders of magnitude faster than traditional sky location methods, taking around 18 ms to localize a single GW sample. Currently, work is being done to extend this method for real noise and with BNS and NSBH sources.
 
Once the sky map for a GW event is produced, it is sent out to the EM community.
However, the optical and near-infrared transient contamination rate can be prohibitively large. ML techniques are a promising method for optimising, automating and significantly reducing the number of optical and near-infrared transients that must be manually vetted.
The expected EM emission of a BNS merger is a short-duration Gamma-Ray Burst (GRB) which is quickly followed by an optical afterglow and an optical and near-infrared kilonova \cite{MetzgerBerger2012}. While the prompt GRB shows highly collimated emission along the line of sight of a jet, the kilonova emission is isotropic. Even if the prompt emission of the GRB is off-axis with respect to observer's line of sight, the afterglow and the kilonova can be observed at later times, albeit at much fainter magnitudes. The contamination rate of optical transients in kilonova searches is estimated to be 1.79 deg$^{-2}$ down to a limiting magnitude of $m_i$=23.5 and $m_z$=22.4 in the $i$- and $z$-band, respectively \cite{Cowperthwaite2018}. Extrapolating out to aLIGO's design detector horizon of $D_L = 200$ Mpc, and effectively increasing the sensitivity and limiting magnitude which are required to observe kilonovae at these distances, this rate is expected to increase by a factor of 4 \cite{Cowperthwaite2018}. Types of astrophysical contaminants that may occur in the same time windows are Type Ia and Type II supernovae, flaring objects such as M-dwarf flares, and nuclear activity in active galactic nuclei (AGN). While these objects typically evolve on time scales which are different than the time scales of GRB afterglows and kilonovae, they may appear in single-subtracted images as possible candidates. Follow-up and vetting of these contaminants is a time-consuming process.

Besides the optical astrophysical contamination, a large number of optical image artifacts occurs as a direct result of image processing. {\em Difference imaging}, or {\em image subtraction}, is the main technique used for the identification of novel transients. Image subtraction is the process of subtracting a new image against an older reference image, removing consistent steady-state brightness sources and ideally leaving behind only novel transients. 
  
Because of the high computational cost of image subtraction, many ML techniques have been employed to improve the rate of image processing contamination.  While there are extensive studies in literature for transient identification with ML, one method directly addresses the reduction of image artifacts in the searches for the EM counterparts of GW signals over their hundreds to thousands of square degree error region skymaps \cite{Ackley2019}. As true transients should appear point-like in the subtracted image, the Point-Spread Function (PSF) of all objects in the subtracted image can be compared directly against point-like PSFs from the convolved reference image. The PSF of the reference image is derived by decomposing each source in the image onto a set of Zernike polynomials and summing the median of the distribution of coefficients for each order. Every source in the subtracted image is decomposed in the same way and the coefficient for each order is compared against the reference PSF, generating a value for how distant the reference PSF and the subtracted source PSF are in statistical space, called the {\em Zernike distance}. By using shape information alone and employing unsupervised methods, the method returns an efficiency of 91.5\% and 92.8\% on sets of artificial source injection tests with archival images from the Dark Energy Camera \cite{Flaugher:2015pxc} and the Palomar Transient Facility \cite {Rau:2009yx,Law:2009ys}, respectively. The reduction in the number of optical image artifacts is over 99.97\% with over 91\% of the simulated true signals being preserved. This is a reduction from hundreds to thousands of artifacts per image to single or double digit objects, making manual vetting of optical candidates feasible over the hundreds of sq.\ degree error regions.

\section{Conclusions}
\label{sec:conclusion}
 
In this paper, we have provided a review of ML applications to GW science. ML is an exciting area of development in the field of multi-messenger astrophysics. Over the years, LIGO and Virgo researchers have applied a variety of ML algorithms to many challenging problems in GW data analysis and detector characterization.  

The data from LIGO and Virgo is non-stationary and non-Gaussian. ML methods can be successfully used to improve the quality of these data. The probability of a GW signal candidate being astrophysical or transient detector noise can be determined by applying ML to the data from thousands of environmental and instrumental monitors. The origin of detector noise can be inferred by applying classification techniques to find different types of transient noise. Citizen science projects provide training data for some of these studies. 

Matched filter searches and Bayesian parameter estimation require a priori knowledge of the theoretical templates of the expected GW signals. Providing these signal predictions with full numerical relativity is computationally expensive. ML can be used to predict the GW waveforms in areas of the signal parameter space not covered by full numerical relativity. Searches based on ML techniques have comparable search sensitivity to matched filter searches. They are promising tools for future GW searches. 

ML methods can also be successfully applied in GW searches where the exact signal morphology is unknown. Various ML algorithms have been developed to increase the search sensitivity of the standard LIGO-Virgo searches for GW bursts, as well as searches for GWs from CCSN and longer duration continuous GW signals.  

After a GW signal has been detected, LIGO and Virgo run computationally expensive parameter estimation codes to determine the characteristics of the source. We have reviewed several studies showing that ML algorithms can significantly speed up parameter estimation of GW signals. After multiple detections are made, ML can be applied to determine the populations of GW sources and their properties, which will inform us of their formation mechanisms. Finally, ML can aid in finding EM counterparts to GW signals. 

Considering the future improvements in the sensitivity of GW detectors, and their ability to detect many events per week, ML techniques are poised to become essential tools in GW science and multi-messenger astrophysics.

\section*{Acknowledgments}

We thank Jess McIver and Damir Buskulic for their feedback on this work. \\

This publication is supported by work from COST Action CA17137, supported by COST (European Cooperation in Science and Technology).
JP, KA and PE are supported by the Australian Research Council Centre of Excellence for Gravitational Wave Discovery (OzGrav), through project number CE170100004.
MC is supported by the National Science Foundation through award PHY-1921006.
LH is supported by the Swiss National Science Foundation with the Early Postdoc Mobility grant number 181461.
DW and HG are supported by Science and Technology Facilities Council (STFC) grant ST/L000946/1. 
RE is supported at the University of Chicago by the Kavli Institute for Cosmological Physics through an endowment from the Kavli Foundation and its founder Fred Kavli.
SM and ZM thank Columbia University in the City of New York for their generous support and are supported by the National Science Foundation under grant CCF-1740391. SM and ZM acknowledge computing resources from Columbia University's Shared Research Computing Facility project, which is supported by NIH Research Facility Improvement Grant 1G20RR030893-01, and associated funds from the New York State Empire State Development, Division of Science Technology and Innovation (NYSTAR) Contract C090171, both awarded April 15, 2010.
TDG acknowledges partial funding from the Max Planck ETH Center for Learning Systems.
Gravity Spy and SC is partly supported by the National Science
Foundation award INSPIRE 15-47880.
VG is supported by the LIGO Laboratory, NSF grant PHY-1764464.
DK is supported by the Spanish Ministry of Science, Innovation and Universities grant FPA2016-76821 and the Vicepresid{\`e}ncia i Conselleria d'Innovaci{\'o}, Recerca i Turisme and Conselleria d'Educaci{\'o} i Universitats of the Govern de les Illes Balears.
MB and FM are partially supported by the Polish National Science Centre grants no. 2016/22/E/ST9/00037 and 2017/26/M/ST9/00978.
LIGO was constructed by the California Institute of Technology and Massachusetts Institute of Technology with funding from the United States National Science Foundation under grant PHY-0757058.
The authors are grateful for computational resources provided by the LIGO Laboratory and supported by the National Science Foundation Grants PHY-0757058 and PHY-0823459.


\bibliographystyle{iopart-num}
\bibliography{bibfile}

\providecommand{\newblock}{}
\begin{thebibliography}{100}
\expandafter\ifx\csname url\endcsname\relax
  \def\url#1{{\tt #1}}\fi
\expandafter\ifx\csname urlprefix\endcsname\relax\def\urlprefix{URL }\fi
\providecommand{\eprint}[2][]{\url{#2}}

\bibitem{aLIGO}
{The LIGO Scientific Collaboration}, {Aasi} J, {Abbott} B~P, {Abbott} R and
  et~al 2015 {\em \cqg\/} {\bf 32} 074001 (\textit{Preprint}
  \eprint{1411.4547})

\bibitem{AdVirgo}
{Acernese} F and et~al 2015 {\em \cqg\/} {\bf 32} 024001 (\textit{Preprint}
  \eprint{1408.3978})

\bibitem{Abbott:2016blz}
Abbott B~P {\em et~al.\/} (LIGO Scientific, Virgo) 2016 {\em Phys. Rev.
  Lett.\/} {\bf 116} 061102 (\textit{Preprint} \eprint{1602.03837})

\bibitem{Abbott:2016nmj}
Abbott B~P {\em et~al.\/} (LIGO Scientific, Virgo) 2016 {\em Phys. Rev.
  Lett.\/} {\bf 116} 241103 (\textit{Preprint} \eprint{1606.04855})

\bibitem{PhysRevX.9.031040}
{Abbott} B~P {\em et~al.\/} (LIGO Scientific Collaboration and Virgo
  Collaboration) 2019 {\em Phys. Rev. X\/} {\bf 9}(3) 031040
  \urlprefix\url{https://link.aps.org/doi/10.1103/PhysRevX.9.031040}

\bibitem{Abbott_2020}
{Abbott} B~P {\em et~al.\/} (LIGO Scientific Collaboration and Virgo
  Collaboration) 2020 {\em Classical and Quantum Gravity\/} {\bf 37} 055002
  \urlprefix\url{https://doi.org/10.1088%2F1361-6382%2Fab685e}

\bibitem{LIGOScientific:2020stg}
 2020  (\textit{Preprint} \eprint{2004.08342})

\bibitem{graceDB}
{GraceDB -- Gravitational-Wave Candidate Event Database}
  https://gracedb.ligo.org/superevents/public/O3/

\bibitem{Somiya:2011np}
Somiya K (KAGRA) 2012 {\em Class. Quant. Grav.\/} {\bf 29} 124007
  (\textit{Preprint} \eprint{1111.7185})

\bibitem{Akutsu:2018axf}
Akutsu T {\em et~al.\/} (KAGRA) 2019 {\em Nat. Astron.\/} {\bf 3} 35--40
  (\textit{Preprint} \eprint{1811.08079})

\bibitem{ligo-india:2011}
{Iyer B, et al} 2011 {LIGO-India} Tech. rep. LIGO Document Control Center
  \urlprefix\url{https://dcc.ligo.org/ligo-M1100296/public}

\bibitem{LIGOScientific:2018jsj}
Abbott B~P {\em et~al.\/} (LIGO Scientific, Virgo) 2019 {\em Astrophys. J.\/}
  {\bf 882} L24 (\textit{Preprint} \eprint{1811.12940})

\bibitem{LIGOScientific:2019fpa}
Abbott B~P {\em et~al.\/} (LIGO Scientific, Virgo) 2019 {\em Phys. Rev.\/} {\bf
  D100} 104036 (\textit{Preprint} \eprint{1903.04467})

\bibitem{Abbott:2019yzh}
Abbott B~P {\em et~al.\/} (LIGO Scientific, Virgo) 2019  (\textit{Preprint}
  \eprint{1908.06060})

\bibitem{Abbott:2019pxc}
Abbott B~P {\em et~al.\/} (LIGO Scientific, Virgo) 2019  (\textit{Preprint}
  \eprint{1908.03584})

\bibitem{Abbott:2019dxx}
Abbott B~P {\em et~al.\/} (LIGO Scientific, Virgo) 2019 {\em Astrophys. J.\/}
  {\bf 874} 163 (\textit{Preprint} \eprint{1902.01557})

\bibitem{LIGOScientific:2019vic}
Abbott B~P {\em et~al.\/} (LIGO Scientific, Virgo) 2019 {\em Phys. Rev.\/} {\bf
  D100} 061101 (\textit{Preprint} \eprint{1903.02886})

\bibitem{wienerbook}
Wiener N 1949 {\em { Extrapolation, Interpolation, and Smoothing of Stationary
  Time Series}\/} (New York: Wiley)

\bibitem{PhysRevD.44.3819}
Sathyaprakash B~S and Dhurandhar S~V 1991 {\em Phys. Rev. D\/} {\bf 44}(12)
  3819--3834 \urlprefix\url{https://link.aps.org/doi/10.1103/PhysRevD.44.3819}

\bibitem{2016PhRvD..93d2004K}
{Klimenko} S {\em et~al.\/} 2016 {\em \prd\/} {\bf 93} 042004
  (\textit{Preprint} \eprint{1511.05999})

\bibitem{0264-9381-25-11-114029}
Klimenko S {\em et~al.\/} 2008 {\em Classical and Quantum Gravity\/} {\bf 25}
  114029 \urlprefix\url{http://stacks.iop.org/0264-9381/25/i=11/a=114029}

\bibitem{PhysRevD.59.102001}
Allen B and Romano J~D 1999 {\em Phys. Rev. D\/} {\bf 59}(10) 102001
  \urlprefix\url{https://link.aps.org/doi/10.1103/PhysRevD.59.102001}

\bibitem{powell:15}
{Powell} J, {Trifir{\`o}} D, {Cuoco} E, {Heng} I~S and {Cavagli{\`a}} M 2015
  {\em \cqg\/} {\bf 32} 215012 (\textit{Preprint} \eprint{1505.01299})

\bibitem{powell:16}
{Powell} J, {Torres-Forn{\'e}} A, {Lynch} R, {Trifir{\`o}} D, {Cuoco} E,
  {Cavagli{\`a}} M, {Heng} I~S and {Font} J~A 2017 {\em Classical and Quantum
  Gravity\/} {\bf 34} 034002 (\textit{Preprint} \eprint{1609.06262})

\bibitem{PhysRevD.95.104059}
Mukund N, Abraham S, Kandhasamy S, Mitra S and Philip N~S 2017 {\em Phys. Rev.
  D\/} {\bf 95}(10) 104059
  \urlprefix\url{https://link.aps.org/doi/10.1103/PhysRevD.95.104059}

\bibitem{PhysRevD.97.101501}
George D, Shen H and Huerta E~A 2018 {\em Phys. Rev. D\/} {\bf 97}(10) 101501
  \urlprefix\url{https://link.aps.org/doi/10.1103/PhysRevD.97.101501}

\bibitem{2013PhRvD..88f2003B}
{Biswas} R {\em et~al.\/} 2013 {\em \prd\/} {\bf 88} 062003 (\textit{Preprint}
  \eprint{1303.6984})

\bibitem{Pinto2013}
Rampone S, Pierro V, Troiano L and Pinto I~M 2013 {\em International Journal of
  Modern Physics C\/} {\bf 24} ISSN 0129-1831

\bibitem{Lightman_2006}
Lightman M, Thurakal J, Dwyer J, Grossman R, Kalmus P, Matone L, Rollins J,
  Zairis S and M{\'{a}}rka S 2006 {\em Journal of Physics: Conference Series\/}
  {\bf 32} 58--65

\bibitem{2018razzano}
Razzano M and Cuoco E 2018 {\em Classical and Quantum Gravity\/} {\bf 35}
  095016 \urlprefix\url{https://doi.org/10.1088%2F1361-6382%2Faab793}

\bibitem{Huerta:2019rtg}
Huerta E~A {\em et~al.\/} 2019 {\em Nature Rev. Phys.\/} {\bf 1} 600--608
  (\textit{Preprint} \eprint{1911.11779})

\bibitem{Cavaglia:2020qzp}
{Cavagli{\`a}, Marco and Gaudio, Sergio and Hansen, Travis and Staats, Kai and
  Szczepanczyk, Marek and Zanolin, Michele} 2020 {\em Sci. Technol.\/} {\bf 1}
  015005 (\textit{Preprint} \eprint{2002.04591})

\bibitem{Kim_2015}
Kim K, Harry I~W, Hodge K~A, Kim Y~M, Lee C~H, Lee H~K, Oh J~J, Oh S~H and Son
  E~J 2015 {\em Classical and Quantum Gravity\/} {\bf 32} 245002 ISSN 1361-6382
  \urlprefix\url{http://dx.doi.org/10.1088/0264-9381/32/24/245002}

\bibitem{Kim_2020}
Kim K, Li T~G, Lo R~K, Sachdev S and Yuen R~S 2020 {\em Physical Review D\/}
  {\bf 101} ISSN 2470-0029
  \urlprefix\url{http://dx.doi.org/10.1103/PhysRevD.101.083006}

\bibitem{PhysRevLett.120.141103}
Gabbard H, Williams M, Hayes F and Messenger C 2018 {\em Phys. Rev. Lett.\/}
  {\bf 120}(14) 141103
  \urlprefix\url{https://link.aps.org/doi/10.1103/PhysRevLett.120.141103}

\bibitem{2018PhLB..778...64G}
{George} D and {Huerta} E~A 2018 {\em Physics Letters B\/} {\bf 778} 64--70
  (\textit{Preprint} \eprint{1711.03121})

\bibitem{2012MNRAS.421..169G}
{Graff} P, {Feroz} F, {Hobson} M~P and {Lasenby} A 2012 {\em \mnras\/} {\bf
  421} 169--180 (\textit{Preprint} \eprint{1110.2997})

\bibitem{2019arXiv190301998S}
{Shen} H, {Huerta} E~A and {Zhao} Z 2019 {\em arXiv e-prints\/}
  arXiv:1903.01998 (\textit{Preprint} \eprint{1903.01998})

\bibitem{1909.06296}
Gabbard H, Messenger C, Heng I~S, Tonolini F and Murray-Smith R 2019 Bayesian
  parameter estimation using conditional variational autoencoders for
  gravitational-wave astronomy (\textit{Preprint} \eprint{arXiv:1909.06296})

\bibitem{PhysRevLett.124.041102}
Chua A~J~K and Vallisneri M 2020 {\em Phys. Rev. Lett.\/} {\bf 124}(4) 041102
  \urlprefix\url{https://link.aps.org/doi/10.1103/PhysRevLett.124.041102}

\bibitem{VaHu2019}
Vajente G, Huang Y, Isi M, Driggers J~C, Kissel J~S,
  Szczepa\ifmmode~\acute{n}\else \'{n}\fi{}czyk M~J and Vitale S 2020 {\em
  Phys. Rev. D\/} {\bf 101}(4) 042003
  \urlprefix\url{https://link.aps.org/doi/10.1103/PhysRevD.101.042003}

\bibitem{gravityspy}
Zevin M, Coughlin S, Bahaadini S, Besler E, Rohani N, Allen S, Cabero M,
  Crowston K, Katsaggelos A~K, Larson S~L, Lee T~K, Lintott C, Littenberg T~B,
  Lundgren A, Østerlund C, Smith J~R, Trouille L and Kalogera V 2017 {\em
  Classical and Quantum Gravity\/} {\bf 34} 064003
  \urlprefix\url{http://stacks.iop.org/0264-9381/34/i=6/a=064003}

\bibitem{AbEA2016g}
{LIGO Scientific Collaboration and Virgo Collaboration} 2016 {\em Classical and
  Quantum Gravity\/} {\bf 33} 134001
  \urlprefix\url{http://stacks.iop.org/0264-9381/33/i=13/a=134001}

\bibitem{BAHAADINI2018172}
Bahaadini S, Noroozi V, Rohani N, Coughlin S, Zevin M, Smith J, Kalogera V and
  Katsaggelos A 2018 {\em Information Sciences\/} {\bf 444} 172 -- 186 ISSN
  0020-0255
  \urlprefix\url{http://www.sciencedirect.com/science/article/pii/S0020025518301634}

\bibitem{Powell:2015ona}
Powell J, Trifir{\`o} D, Cuoco E, Heng I~S and Cavagli{\`a} M 2015 {\em Class.
  Quant. Grav.\/} {\bf 32} 215012 (\textit{Preprint} \eprint{1505.01299})

\bibitem{Powell:2016rkl}
Powell J, Torres-Forn{\'e} A, Lynch R, Trifir{\`o} D, Cuoco E, Cavagli{\`a} M,
  Heng I~S and Font J~A 2017 {\em Class. Quant. Grav.\/} {\bf 34} 034002
  (\textit{Preprint} \eprint{1609.06262})

\bibitem{2017Mukund}
{Mukund} N, {Abraham} S, {Kandhasamy} S, {Mitra} S and {Philip} N~S 2017 {\em
  \prd\/} {\bf 95} 104059 (\textit{Preprint} \eprint{1609.07259})

\bibitem{Yamashita2018}
Yamashita R, Nishio M, Do R~K~G and Togashi K 2018 {\em Insights into
  Imaging\/} {\bf 9} 611--629

\bibitem{Russakovsky2015}
Russakovsky O {\em et~al.\/} 2015 {\em Int. J. Comput. Vis.\/} {\bf 115}
  211--252

\bibitem{rollinsthesis}
{Rollins} J 2011 {\em {Multimessenger Astronomy with Low-Latency Searches for
  Transient Gravitational Waves}\/} Ph.D. thesis {Columbia University}

\bibitem{2004CQGra..21S1809C}
{Chatterji} S, {Blackburn} L, {Martin} G and {Katsavounidis} E 2004 {\em
  Classical and Quantum Gravity\/} {\bf 21} S1809--S1818 (\textit{Preprint}
  \eprint{gr-qc/0412119})

\bibitem{8553393}
{Cuoco} E, {Razzano} M and {Utina} A 2018 Wavelet-based classification of
  transient signals for gravitational wave detectors {\em 2018 26th European
  Signal Processing Conference (EUSIPCO)\/} pp 2648--2652 ISSN 2219-5491

\bibitem{WDF-GRB}
Acernese F {\em et~al.\/} 2007 {\em Classical and Quantum Gravity\/} {\bf 24}
  S671 \urlprefix\url{http://stacks.iop.org/0264-9381/24/i=19/a=S29}

\bibitem{XGBoost}
Tianqi~Chen C~G 2016 Xgboost: A scalable tree boosting system. {\em KDD '16
  Proceedings of the 22nd ACM SIGKDD International Conference on Knowledge
  Discovery and Data Mining\/} pp 785--794

\bibitem{2013CQGra..30o5010E}
{Essick} R, {Blackburn} L and {Katsavounidis} E 2013 {\em Classical and Quantum
  Gravity\/} {\bf 30} 155010 (\textit{Preprint} \eprint{1303.7159})

\bibitem{2011CQGra..28w5005S}
{Smith} J~R, {Abbott} T, {Hirose} E, {Leroy} N, {MacLeod} D, {McIver} J,
  {Saulson} P and {Shawhan} P 2011 {\em Classical and Quantum Gravity\/} {\bf
  28} 235005 (\textit{Preprint} \eprint{1107.2948})

\bibitem{2010JPhCS.243a2005I}
{Isogai} T, {LIGO Scientific Collaboration} and {Virgo Collaboration} 2010
  {Used percentage veto for LIGO and virgo binary inspiral searches} {\em
  Journal of Physics Conference Series\/} vol 243 p 012005

\bibitem{Colgan:2019lyo}
Colgan R~E, Corley K~R, Lau Y, Bartos I, Wright J~N, Marka Z and Marka S 2019
  (\textit{Preprint} \eprint{1911.11831})

\bibitem{essick_thesis}
Essick R 2017 {\em Detectability of dynamical tidal effects and the detection
  of gravitational-wave transients with LIGO\/} Ph.D. thesis Massachusetts
  Institute of Technology

\bibitem{Essick2020}
Essick R {\em et~al.\/} 2020 idq: Statistical inference for non-gaussian noise
  with auxiliary degrees of freedom in gravitational-wave detectors in prep.

\bibitem{Godwin2020}
Godwin P {\em et~al.\/} 2020 Stream-based noise acquisition and extraction
  analysis (snax) in prep.

\bibitem{PhysRevLett.119.161101}
Abbott B~P {\em et~al.\/} (LIGO Scientific Collaboration and Virgo
  Collaboration) 2017 {\em Phys. Rev. Lett.\/} {\bf 119}(16) 161101
  \urlprefix\url{https://link.aps.org/doi/10.1103/PhysRevLett.119.161101}

\bibitem{Cavaglia:2018xjq}
Cavagli{\`a} M, Staats K and Gill T 2019 {\em Commun. Comput. Phys.\/} {\bf 25}
  963--987 (\textit{Preprint} \eprint{1812.05225})

\bibitem{Omicron}
Robinet F 2015  \urlprefix\url{https://tds.ego-gw.it/ql/?c=10651}

\bibitem{Bishop}
Bishop C~M 2006 {\em Pattern Recognition and Machine Learning\/} (Berlin,
  Heidelberg: Springer-Verlag) ISBN 0387310738

\bibitem{Hastie}
Hastie T, Tibshirani R and Friedman J 2009 {\em The Elements of Statistical
  Learning: Data Mining, Inference and Prediction\/} 2nd ed (Springer)
  \urlprefix\url{http://www-stat.stanford.edu/~tibs/ElemStatLearn/}

\bibitem{2017CQGra..34f4003Z}
{Zevin} M, {Coughlin} S, {Bahaadini} S, {Besler} E, {Rohani} N, {Allen} S,
  {Cabero} M, {Crowston} K, {Katsaggelos} A~K, {Larson} S~L, {Lee} T~K,
  {Lintott} C, {Littenberg} T~B, {Lundgren} A, {{\O}sterlund} C, {Smith} J~R,
  {Trouille} L and {Kalogera} V 2017 {\em Classical and Quantum Gravity\/} {\bf
  34} 064003 (\textit{Preprint} \eprint{1611.04596})

\bibitem{Mukund:2018occ}
Mukund N {\em et~al.\/} 2019 {\em Class.\ Quant.\ Grav.\/} {\bf 36} 085005
  (\textit{Preprint} \eprint{1812.05185})

\bibitem{Mukund:2017fal}
Mukund N, Thakur S, Abraham S, Aniyan A, Mitra S, Philip N~S, Vaghmare K and
  Acharjya D 2018 {\em Astrophys.\ J.\ Suppl.\/} {\bf 235} 22
  (\textit{Preprint} \eprint{1710.05350})

\bibitem{DeepClean}
et~al O 2020 {\em In preparation.\/}

\bibitem{2014PhRvD..90h4029T}
{Torres} A, {Marquina} A, {Font} J~A and {Ib{\'a}{\~n}ez} J~M 2014 {\em \prd\/}
  {\bf 90} 084029 (\textit{Preprint} \eprint{1409.7888})

\bibitem{2018PhRvD..98h4013T}
{Torres-Forn{\'e}} A, {Cuoco} E, {Marquina} A, {Font} J~A and {Ib{\'a}{\~n}ez}
  J~M 2018 {\em \prd\/} {\bf 98} 084013 (\textit{Preprint} \eprint{1806.07329})

\bibitem{2015ASSP...40..289T}
{Torres} A, {Marquina} A, {Font} J~A and {Ib{\'a}{\~n}ez} J~M 2015 {Split
  Bregman Method for Gravitational Wave Denoising} {\em Gravitational Wave
  Astrophysics\/} vol~40 p 289

\bibitem{2016PhRvD..94l4040T}
{Torres-Forn{\'e}} A, {Marquina} A, {Font} J~A and {Ib{\'a}{\~n}ez} J~M 2016
  {\em \prd\/} {\bf 94} 124040 (\textit{Preprint} \eprint{1612.01305})

\bibitem{alej2020application}
Torres-Forné A, Cuoco E, Font J~A and Marquina A 2020 Application of
  dictionary learning to denoise ligo's blip noise transients
  (\textit{Preprint} \eprint{2002.11668})

\bibitem{2020PhLB..80035081W}
{Wei} W and {Huerta} E~A 2020 {\em Physics Letters B\/} {\bf 800} 135081
  (\textit{Preprint} \eprint{1901.00869})

\bibitem{2016arXiv160903499V}
{van den Oord} A, {Dieleman} S, {Zen} H, {Simonyan} K, {Vinyals} O, {Graves} A,
  {Kalchbrenner} N, {Senior} A and {Kavukcuoglu} K 2016 {\em arXiv e-prints\/}
  arXiv:1609.03499 (\textit{Preprint} \eprint{1609.03499})

\bibitem{10.5555/553011}
Jain L~C and Medsker L~R 1999 {\em Recurrent Neural Networks: Design and
  Applications\/} 1st ed (USA: CRC Press, Inc.) ISBN 0849371813

\bibitem{2013arXiv1312.6026P}
{Pascanu} R, {Gulcehre} C, {Cho} K and {Bengio} Y 2013 {\em arXiv e-prints\/}
  arXiv:1312.6026 (\textit{Preprint} \eprint{1312.6026})

\bibitem{2018APS..APRS14008S}
{Shen} H, {Zhao} Z, {George} D and {Huerta} E 2018 {Denoising Gravitational
  Waves using Deep Learning with Recurrent Denoising Autoencoders} {\em APS
  April Meeting Abstracts\/} ({\em APS Meeting Abstracts\/} vol 2018) p S14.008

\bibitem{2019arXiv190303105S}
{Shen} H, {George} D, {Huerta} E~A and {Zhao} Z 2019 {\em arXiv e-prints\/}
  arXiv:1903.03105 (\textit{Preprint} \eprint{1903.03105})

\bibitem{Nitz:2018rgo}
Nitz A~H, Dal~Canton T, Davis D and Reyes S 2018 {\em Phys. Rev. D\/} {\bf 98}
  024050 (\textit{Preprint} \eprint{1805.11174})

\bibitem{Sachdev:2019vvd}
Sachdev S {\em et~al.\/} 2019  (\textit{Preprint} \eprint{1901.08580})

\bibitem{Messick:2016aqy}
Messick C {\em et~al.\/} 2017 {\em Phys. Rev.\/} {\bf D95} 042001
  (\textit{Preprint} \eprint{1604.04324})

\bibitem{Adams_2016}
Adams T, Buskulic D, Germain V, Guidi G~M, Marion F, Montani M, Mours B,
  Piergiovanni F and Wang G 2016 {\em Classical and Quantum Gravity\/} {\bf 33}
  175012 \urlprefix\url{https://doi.org/10.1088%2F0264-9381%2F33%2F17%2F175012}

\bibitem{SPIIR}
Chu Q 2017 Ph.D. thesis The University of Western Australia

\bibitem{veitch:15}
{Veitch} J, {Raymond} V, {Farr} B, {Farr} W, {Graff} P, {Vitale} S, {Aylott} B,
  {Blackburn} K, {Christensen} N, {Coughlin} M, {Del Pozzo} W, {Feroz} F,
  {Gair} J, {Haster} C~J, {Kalogera} V, {Littenberg} T, {Mandel} I,
  {O'Shaughnessy} R, {Pitkin} M, {Rodriguez} C, {R{\"o}ver} C, {Sidery} T,
  {Smith} R, {Van Der Sluys} M, {Vecchio} A, {Vousden} W and {Wade} L 2015 {\em
  \prd\/} {\bf 91} 042003 (\textit{Preprint} \eprint{1409.7215})

\bibitem{2019ApJS..241...27A}
{Ashton} G, {H{\"u}bner} M, {Lasky} P~D, {Talbot} C, {Ackley} K, {Biscoveanu}
  S, {Chu} Q, {Divakarla} A, {Easter} P~J and {Goncharov} B 2019 {\em \apjs\/}
  {\bf 241} 27 (\textit{Preprint} \eprint{1811.02042})

\bibitem{Pretorius:2005gq}
Pretorius F 2005 {\em Phys. Rev. Lett.\/} {\bf 95} 121101 (\textit{Preprint}
  \eprint{gr-qc/0507014})

\bibitem{Campanelli:2005dd}
Campanelli M, Lousto C~O, Marronetti P and Zlochower Y 2006 {\em Phys. Rev.
  Lett.\/} {\bf 96} 111101 (\textit{Preprint} \eprint{gr-qc/0511048})

\bibitem{Baker:2005vv}
Baker J~G, Centrella J, Choi D~I, Koppitz M and van Meter J 2006 {\em Phys.
  Rev. Lett.\/} {\bf 96} 111102 (\textit{Preprint} \eprint{gr-qc/0511103})

\bibitem{Bruegmann:2006at}
Bruegmann B, Gonzalez J~A, Hannam M, Husa S, Sperhake U and Tichy W 2008 {\em
  Phys. Rev.\/} {\bf D77} 024027 (\textit{Preprint} \eprint{gr-qc/0610128})

\bibitem{Centrella:2010mx}
Centrella J, Baker J~G, Kelly B~J and van Meter J~R 2010 {\em Rev.Mod.Phys.\/}
  {\bf 82} 3069 (\textit{Preprint} \eprint{1010.5260})

\bibitem{Mroue:2013xna}
Mroue A~H {\em et~al.\/} 2013 {\em Phys. Rev. Lett.\/} {\bf 111} 241104
  (\textit{Preprint} \eprint{1304.6077})

\bibitem{Jani:2016wkt}
Jani K, Healy J, Clark J~A, London L, Laguna P and Shoemaker D 2016 {\em Class.
  Quant. Grav.\/} {\bf 33} 204001 (\textit{Preprint} \eprint{1605.03204})

\bibitem{Healy:2017psd}
Healy J, Lousto C~O, Zlochower Y and Campanelli M 2017 {\em Class. Quant.
  Grav.\/} {\bf 34} 224001 (\textit{Preprint} \eprint{1703.03423})

\bibitem{Boyle:2019kee}
Boyle M {\em et~al.\/} 2019 {\em Class. Quant. Grav.\/} {\bf 36} 195006
  (\textit{Preprint} \eprint{1904.04831})

\bibitem{Blanchet:2013haa}
Blanchet L 2014 {\em Living Rev. Rel.\/} {\bf 17} 2 (\textit{Preprint}
  \eprint{1310.1528})

\bibitem{Berti:2009kk}
Berti E, Cardoso V and Starinets A~O 2009 {\em Class. Quant. Grav.\/} {\bf 26}
  163001 (\textit{Preprint} \eprint{0905.2975})

\bibitem{Buonanno:1998gg}
Buonanno A and Damour T 1999 {\em Phys. Rev.\/} {\bf D59} 084006
  (\textit{Preprint} \eprint{gr-qc/9811091})

\bibitem{Buonanno:2000ef}
Buonanno A and Damour T 2000 {\em Phys. Rev.\/} {\bf D62} 064015

\bibitem{Damour:2000we}
Damour T, Jaranowski P and Schaefer G 2000 {\em Phys.Rev.\/} {\bf D62} 084011
  (\textit{Preprint} \eprint{gr-qc/0005034})

\bibitem{Damour:2001tu}
Damour T 2001 {\em Phys.Rev.\/} {\bf D64} 124013 (\textit{Preprint}
  \eprint{gr-qc/0103018})

\bibitem{Damour:2008gu}
Damour T, Iyer B~R and Nagar A 2009 {\em Phys.Rev.\/} {\bf D79} 064004
  (\textit{Preprint} \eprint{0811.2069})

\bibitem{Pan:2011gk}
Pan Y, Buonanno A, Boyle M, Buchman L~T, Kidder L~E, Pfeiffer H~P and Scheel
  M~A 2011 {\em Phys. Rev.\/} {\bf D84} 124052 (\textit{Preprint}
  \eprint{1106.1021})

\bibitem{Taracchini:2012ig}
Taracchini A, Pan Y, Buonanno A, Barausse E, Boyle M {\em et~al.\/} 2012 {\em
  Phys.Rev.\/} {\bf D86} 024011 (\textit{Preprint} \eprint{1202.0790})

\bibitem{Taracchini:2013rva}
Taracchini A, Buonanno A, Pan Y, Hinderer T, Boyle M {\em et~al.\/} 2014 {\em
  Phys.Rev.\/} {\bf D89} 061502 (\textit{Preprint} \eprint{1311.2544})

\bibitem{Pan:2013rra}
Pan Y, Buonanno A, Taracchini A, Kidder L~E, Mrou\'e A~H, Pfeiffer H~P, Scheel
  M~A and Szil\'agyi B 2014 {\em Phys. Rev.\/} {\bf D89} 084006
  (\textit{Preprint} \eprint{1307.6232})

\bibitem{Damour:2014sva}
Damour T and Nagar A 2014 {\em Phys. Rev.\/} {\bf D90} 044018
  (\textit{Preprint} \eprint{1406.6913})

\bibitem{Nagar:2015xqa}
Nagar A, Damour T, Reisswig C and Pollney D 2016 {\em Phys. Rev.\/} {\bf D93}
  044046 (\textit{Preprint} \eprint{1506.08457})

\bibitem{Bohe:2016gbl}
Boh\'e A {\em et~al.\/} 2017 {\em Phys. Rev.\/} {\bf D95} 044028
  (\textit{Preprint} \eprint{1611.03703})

\bibitem{Babak:2016tgq}
Babak S, Taracchini A and Buonanno A 2017 {\em Phys. Rev.\/} {\bf D95} 024010
  (\textit{Preprint} \eprint{1607.05661})

\bibitem{Knowles:2018hqq}
Knowles T~D, Devine C, Buch D~A, Bilgili S~A, Adams T~R, Etienne Z~B and
  Mcwilliams S~T 2018 {\em Class. Quant. Grav.\/} {\bf 35} 155003
  (\textit{Preprint} \eprint{1803.06346})

\bibitem{Cotesta:2018fcv}
Cotesta R, Buonanno A, Boh\'e A, Taracchini A, Hinder I and Ossokine S 2018
  {\em Phys. Rev.\/} {\bf D98} 084028 (\textit{Preprint} \eprint{1803.10701})

\bibitem{Nagar:2018zoe}
Nagar A {\em et~al.\/} 2018 {\em Phys. Rev.\/} {\bf D98} 104052
  (\textit{Preprint} \eprint{1806.01772})

\bibitem{Ajith:2011}
Ajith P {\em et~al.\/} 2011 {\em Phys. Rev. Lett.\/} {\bf 106} 241101
  (\textit{Preprint} \eprint{0909.2867})

\bibitem{Santamaria:2010yb}
Santamar{\'\i}a L {\em et~al.\/} 2010  {\bf 82} 064016 (\textit{Preprint}
  \eprint{1005.3306})

\bibitem{Hannam:2013oca}
{Hannam} M, {Schmidt} P, {Boh{\'e}} A, {Haegel} L, {Husa} S, {Ohme} F,
  {Pratten} G and {P{\"u}rrer} M 2014 {\em \prl\/} {\bf 113} 151101
  (\textit{Preprint} \eprint{1308.3271})

\bibitem{PhenomPv2_tech_rep}
Boh\'e A, Hannam M, Husa S, Ohme F, P{\"u}rrer M and Schmidt P 2016 {PhenomPv2
  - Technical notes for the LAL implementation} Tech. rep. LIGO Document
  Control Center \urlprefix\url{https://dcc.ligo.org/LIGO-T1500602/public}

\bibitem{Husa:2015iqa}
Husa S, Khan S, Hannam M, P{\"u}rrer M, Ohme F, Jim\'enez~Forteza X and Boh\'e
  A 2016 {\em Phys. Rev.\/} {\bf D93} 044006 (\textit{Preprint}
  \eprint{1508.07250})

\bibitem{Khan:2015jqa}
Khan S, Husa S, Hannam M, Ohme F, P{\"u}rrer M, Jim\'enez~Forteza X and Boh\'e
  A 2016 {\em Phys. Rev.\/} {\bf D93} 044007 (\textit{Preprint}
  \eprint{1508.07253})

\bibitem{Mehta:2017jpq}
Mehta A~K, Mishra C~K, Varma V and Ajith P 2017 {\em Phys. Rev.\/} {\bf D96}
  124010 (\textit{Preprint} \eprint{1708.03501})

\bibitem{London:2017bcn}
London L, Khan S, Fauchon-Jones E, Garc{\'\i}a C, Hannam M, Husa S,
  Jim\'enez-Forteza X, Kalaghatgi C, Ohme F and Pannarale F 2018 {\em Phys.
  Rev. Lett.\/} {\bf 120} 161102 (\textit{Preprint} \eprint{1708.00404})

\bibitem{Khan:2018fmp}
Khan S, Chatziioannou K, Hannam M and Ohme F 2019 {\em Phys. Rev.\/} {\bf D100}
  024059 (\textit{Preprint} \eprint{1809.10113})

\bibitem{Khan:2019kot}
Khan S, Ohme F, Chatziioannou K and Hannam M 2020 {\em Phys. Rev.\/} {\bf D101}
  024056 (\textit{Preprint} \eprint{1911.06050})

\bibitem{Pratten:2020fqn}
Pratten G, Husa S, Garcia-Quiros C, Colleoni M, Ramos-Buades A, Estelles H and
  Jaume R 2020  (\textit{Preprint} \eprint{2001.11412})

\bibitem{Field:2013cfa}
Field S~E, Galley C~R, Hesthaven J~S, Kaye J and Tiglio M 2014 {\em Phys.
  Rev.\/} {\bf X4} 031006 (\textit{Preprint} \eprint{1308.3565})

\bibitem{Purrer:2014fza}
P{\"u}rrer M 2014 {\em Class. Quant. Grav.\/} {\bf 31} 195010
  (\textit{Preprint} \eprint{1402.4146})

\bibitem{Purrer:2015tud}
P{\"u}rrer M 2016 {\em Phys. Rev.\/} {\bf D93} 064041 (\textit{Preprint}
  \eprint{1512.02248})

\bibitem{Blackman:2015pia}
Blackman J, Field S~E, Galley C~R, Szil\'agyi B, Scheel M~A, Tiglio M and
  Hemberger D~A 2015 {\em Phys. Rev. Lett.\/} {\bf 115} 121102
  (\textit{Preprint} \eprint{1502.07758})

\bibitem{Blackman:2017pcm}
Blackman J, Field S~E, Scheel M~A, Galley C~R, Ott C~D, Boyle M, Kidder L~E,
  Pfeiffer H~P and Szil\'agyi B 2017 {\em Phys. Rev.\/} {\bf D96} 024058
  (\textit{Preprint} \eprint{1705.07089})

\bibitem{Blackman:2017dfb}
Blackman J, Field S~E, Scheel M~A, Galley C~R, Hemberger D~A, Schmidt P and
  Smith R 2017 {\em Phys. Rev.\/} {\bf D95} 104023 (\textit{Preprint}
  \eprint{1701.00550})

\bibitem{Lackey:2018zvw}
Lackey B~D, P{\"u}rrer M, Taracchini A and Marsat S 2019 {\em Phys. Rev.\/}
  {\bf D100} 024002 (\textit{Preprint} \eprint{1812.08643})

\bibitem{Doctor:2017csx}
Doctor Z, Farr B, Holz D~E and P{\"u}rrer M 2017 {\em Phys. Rev.\/} {\bf D96}
  123011 (\textit{Preprint} \eprint{1706.05408})

\bibitem{Varma:2018mmi}
Varma V, Field S~E, Scheel M~A, Blackman J, Kidder L~E and Pfeiffer H~P 2019
  {\em Phys. Rev.\/} {\bf D99} 064045 (\textit{Preprint} \eprint{1812.07865})

\bibitem{Varma:2019csw}
Varma V, Field S~E, Scheel M~A, Blackman J, Gerosa D, Stein L~C, Kidder L~E and
  Pfeiffer H~P 2019 {\em Phys. Rev. Research.\/} {\bf 1} 033015
  (\textit{Preprint} \eprint{1905.09300})

\bibitem{Williams:2019vub}
Williams D, Heng I~S, Gair J, Clark J~A and Khamesra B 2019  (\textit{Preprint}
  \eprint{1903.09204})

\bibitem{williams2019precessing}
Williams D, Heng I~S, Gair J, Clark J~A and Khamesra B 2019 A precessing
  numerical relativity waveform surrogate model for binary black holes: A
  gaussian process regression approach (\textit{Preprint} \eprint{1903.09204})

\bibitem{Moore:2014pda}
Moore C~J and Gair J~R 2014 {\em Phys. Rev. Lett.\/} {\bf 113} 251101
  (\textit{Preprint} \eprint{1412.3657})

\bibitem{Moore:2015sza}
Moore C~J, Berry C~P~L, Chua A~J~K and Gair J~R 2016 {\em Phys. Rev.\/} {\bf
  D93} 064001 (\textit{Preprint} \eprint{1509.04066})

\bibitem{Setyawati:2019xzw}
{Setyawati, Yoshinta and P{\"u}rrer, Michael and Ohme, Frank} 2020 {\em Class.
  Quant. Grav.\/} {\bf 37} 075012 (\textit{Preprint} \eprint{1909.10986})

\bibitem{Setyawati_2020}
Setyawati Y, P{\"u}rrer M and Ohme F 2020 {\em Classical and Quantum Gravity\/}
  {\bf 37} 075012 \urlprefix\url{https://doi.org/10.1088%2F1361-6382%2Fab693b}

\bibitem{Jimenez-Forteza:2016oae}
Jim{\'e}nez-Forteza X, Keitel D, Husa S, Hannam M, Khan S and P{\"u}rrer M 2017
  {\em \prd\/} {\bf 95} 064024 (\textit{Preprint} \eprint{1611.00332})

\bibitem{Healy:2016lce}
Healy J and Lousto C~O 2017 {\em Phys. Rev.\/} {\bf D95} 024037
  (\textit{Preprint} \eprint{1610.09713})

\bibitem{Hofmann:2016yih}
Hofmann F, Barausse E and Rezzolla L 2016 {\em Astrophys. J.\/} {\bf 825} L19
  (\textit{Preprint} \eprint{1605.01938})

\bibitem{Varma:2018aht}
Varma V, Gerosa D, Stein L~C, H\'ebert F and Zhang H 2019 {\em Phys. Rev.
  Lett.\/} {\bf 122} 011101 (\textit{Preprint} \eprint{1809.09125})

\bibitem{Haegel:2019uop}
Haegel L and Husa S 2019  (\textit{Preprint} \eprint{1911.01496})

\bibitem{Easter2019}
Easter P~J, Lasky P~D, Casey A~R, Rezzolla L and Takami K 2019 {\em Physical
  Review D\/} {\bf 100}(4) 043005
  \urlprefix\url{https://link.aps.org/doi/10.1103/PhysRevD.100.043005}

\bibitem{PhysRevD.71.062001}
Allen B 2005 {\em Phys. Rev. D\/} {\bf 71}(6) 062001
  \urlprefix\url{https://link.aps.org/doi/10.1103/PhysRevD.71.062001}

\bibitem{2015PhRvD..91f2004B}
{Baker} P~T, {Caudill} S, {Hodge} K~A, {Talukder} D, {Capano} C and {Cornish}
  N~J 2015 {\em \prd\/} {\bf 91} 062004 (\textit{Preprint} \eprint{1412.6479})

\bibitem{2017PhRvD..96j4015K}
{Kapadia} S~J, {Dent} T and {Dal Canton} T 2017 {\em \prd\/} {\bf 96} 104015
  (\textit{Preprint} \eprint{1709.02421})

\bibitem{George:2017pmj}
George D and Huerta E 2018 {\em Phys. Lett. B\/} {\bf 778} 64--70
  (\textit{Preprint} \eprint{1711.03121})

\bibitem{Gebhard:2019ldz}
Gebhard T~D, Kilbertus N, Harry I and Sch{\"o}lkopf B 2019 {\em Phys. Rev.\/}
  {\bf D100} 063015 (\textit{Preprint} \eprint{1904.08693})

\bibitem{Fryer2003}
Fryer C~L and New K~C~B 2003 {\em Living Reviews in Relativity\/} {\bf 6} 2
  ISSN 1433-8351 \urlprefix\url{https://doi.org/10.12942/lrr-2003-2}

\bibitem{PhysRevLett.87.241101}
Andersson N and Comer G~L 2001 {\em Phys. Rev. Lett.\/} {\bf 87}(24) 241101
  \urlprefix\url{https://link.aps.org/doi/10.1103/PhysRevLett.87.241101}

\bibitem{Baiotti_2007}
Baiotti L, Hawke I and Rezzolla L 2007 {\em Classical and Quantum Gravity\/}
  {\bf 24} S187--S206
  \urlprefix\url{https://doi.org/10.1088%2F0264-9381%2F24%2F12%2Fs13}

\bibitem{PhysRevD.71.063510}
Damour T and Vilenkin A 2005 {\em Phys. Rev. D\/} {\bf 71}(6) 063510
  \urlprefix\url{https://link.aps.org/doi/10.1103/PhysRevD.71.063510}

\bibitem{2017PhRvD..95d2003A}
{Abbott} B~P {\em et~al.\/} (LIGO Scientific Collaboration and Virgo
  Collaboration) 2017 {\em \prd\/} {\bf 95} 042003 (\textit{Preprint}
  \eprint{1611.02972})

\bibitem{2017CQGra..34i4003V}
{Vinciguerra} S, {Drago} M, {Prodi} G~A, {Klimenko} S, {Lazzaro} C, {Necula} V,
  {Salemi} F, {Tiwari} V, {Tringali} M~C and {Vedovato} G 2017 {\em Classical
  and Quantum Gravity\/} {\bf 34} 094003 (\textit{Preprint}
  \eprint{1702.03208})

\bibitem{2019MNRAS.487.1178P}
{Powell} J and {M{\"u}ller} B 2019 {\em \mnras\/} {\bf 487} 1178--1190
  (\textit{Preprint} \eprint{1812.05738})

\bibitem{2019ApJ...876L...9R}
{Radice} D, {Morozova} V, {Burrows} A, {Vartanyan} D and {Nagakura} H 2019 {\em
  \apj\/} {\bf 876} L9 (\textit{Preprint} \eprint{1812.07703})

\bibitem{2019MNRAS.486.2238A}
{Andresen} H, {M{\"u}ller} E, {Janka} H~T, {Summa} A, {Gill} K and {Zanolin} M
  2019 {\em \mnras\/} {\bf 486} 2238--2253 (\textit{Preprint}
  \eprint{1810.07638})

\bibitem{2018MNRAS.475L..91T}
{Takiwaki} T and {Kotake} K 2018 {\em \mnras\/} {\bf 475} L91--L95
  (\textit{Preprint} \eprint{1711.01905})

\bibitem{2019MNRAS.482.3967T}
{Torres-Forn{\'e}} A, {Cerd{\'a}-Dur{\'a}n} P, {Passamonti} A, {Obergaulinger}
  M and {Font} J~A 2019 {\em \mnras\/} {\bf 482} 3967--3988 (\textit{Preprint}
  \eprint{1806.11366})

\bibitem{2018PhRvD..98l2002A}
{Astone} P, {Cerd{\'a}-Dur{\'a}n} P, {Di Palma} I, {Drago} M, {Muciaccia} F,
  {Palomba} C and {Ricci} F 2018 {\em \prd\/} {\bf 98} 122002
  (\textit{Preprint} \eprint{1812.05363})

\bibitem{10.1088/2632-2153/ab7d31}
Iess A, Cuoco E, Morawski F and Powell J 2020 {\em Mach. Learn.: Sci.
  Technol.\/}

\bibitem{Cuoco_2001}
Cuoco E, Calamai G, Fabbroni L, Losurdo G, Mazzoni M, Stanga R and Vetrano F
  2001 {\em Classical and Quantum Gravity\/} {\bf 18} 1727--1751
  \urlprefix\url{https://doi.org/10.1088%2F0264-9381%2F18%2F9%2F309}

\bibitem{2019arXiv191213517C}
{Chan} M~L, {Heng} I~S and {Messenger} C 2019 {\em arXiv e-prints\/}
  arXiv:1912.13517 (\textit{Preprint} \eprint{1912.13517})

\bibitem{Prix:2009oha}
Prix R (for the LIGO Scientific Collaboration) 2009 {\em {Gravitational Waves
  from Spinning Neutron Stars}\/} vol 357 (Springer Berlin Heidelberg) chap~24,
  pp 651--685 ISBN 978-3-540-76964-4
  \urlprefix\url{https://dcc.ligo.org/LIGO-P060039/public}

\bibitem{Brady:1998nj}
Brady P~R and Creighton T 2000  {\bf 61} 082001 (\textit{Preprint}
  \eprint{gr-qc/9812014})

\bibitem{Riles:2017evm}
Riles K 2017 {\em Mod. Phys. Lett.\/} {\bf A32} 1730035 (\textit{Preprint}
  \eprint{1712.05897})

\bibitem{Ming:2017anf}
Ming J, Papa M~A, Krishnan B, Prix R, Beer C, Zhu S~J, Eggenstein H~B, Bock O
  and Machenschalk B 2018 {\em Phys. Rev.\/} {\bf D97} 024051
  (\textit{Preprint} \eprint{1708.02173})

\bibitem{Walsh:2019nmr}
Walsh S, Wette K, Papa M~A and Prix R 2019 {\em Phys. Rev.\/} {\bf D99} 082004
  (\textit{Preprint} \eprint{1901.08998})

\bibitem{Morawski2019DeepLC}
Morawski F, Bejger M and Ciecielkag P 2019 Deep learning classification of the
  continuous gravitational-wave signal candidates from the time-domain
  f-statistic search (\textit{Preprint} \eprint{arXiv:1907.06917})

\bibitem{polgraw-allsky}
Time-domain $\mathcal{F}$-statistic pipeline repository
  \url{https://github.com/mbejger/polgraw-allsky} accessed: 2018-09-04

\bibitem{polgraw-allsky-docs}
Time-domain $\mathcal{F}$-statistic pipeline documentation
  \url{http://mbejger.github.io/polgraw-allsky} accessed: 2018-09-04

\bibitem{Covas:2018oik}
Covas P~B {\em et~al.\/} (LSC) 2018 {\em Phys. Rev.\/} {\bf D97} 082002
  (\textit{Preprint} \eprint{1801.07204})

\bibitem{Pletsch:2008gc}
Pletsch H~J 2008 {\em Phys. Rev.\/} {\bf D78} 102005 (\textit{Preprint}
  \eprint{0807.1324})

\bibitem{Pletsch:2009uu}
Pletsch H~J and Allen B 2009 {\em Phys. Rev. Lett.\/} {\bf 103} 181102
  (\textit{Preprint} \eprint{0906.0023})

\bibitem{Pletsch:2010xb}
Pletsch H~J 2010 {\em Phys. Rev.\/} {\bf D82} 042002 (\textit{Preprint}
  \eprint{1005.0395})

\bibitem{EatH}
{Allen} B {\em et~al.\/} {Einstein@{H}ome distributed computing project}
  https://einsteinathome.org \urlprefix\url{https://einsteinathome.org}

\bibitem{Beheshtipour:2020zhb}
Beheshtipour B and Papa M~A 2020  (\textit{Preprint} \eprint{2001.03116})

\bibitem{Behnke:2014tma}
Behnke B, Papa M~A and Prix R 2015 {\em Phys. Rev.\/} {\bf D91} 064007
  (\textit{Preprint} \eprint{1410.5997})

\bibitem{Papa:2016cwb}
Papa M~A {\em et~al.\/} 2016 {\em Phys. Rev.\/} {\bf D94} 122006
  (\textit{Preprint} \eprint{1608.08928})

\bibitem{Singh:2017kss}
Singh A, Papa M~A, Eggenstein H~B and Walsh S 2017 {\em Phys. Rev.\/} {\bf D96}
  082003 (\textit{Preprint} \eprint{1707.02676})

\bibitem{2013arXiv1311.2524G}
{Girshick} R, {Donahue} J, {Darrell} T and {Malik} J 2013 {\em arXiv
  e-prints\/} arXiv:1311.2524 (\textit{Preprint} \eprint{1311.2524})

\bibitem{2017arXiv170306870H}
{He} K, {Gkioxari} G, {Doll{\'a}r} P and {Girshick} R 2017 {\em arXiv
  e-prints\/} arXiv:1703.06870 (\textit{Preprint} \eprint{1703.06870})

\bibitem{Abbott:2017pqa}
Abbott B~P {\em et~al.\/} (LIGO Scientific, Virgo) 2017 {\em Phys. Rev.\/} {\bf
  D96} 122004 (\textit{Preprint} \eprint{1707.02669})

\bibitem{Dreissigacker:2019edy}
Dreissigacker C, Sharma R, Messenger C, Zhao R and Prix R 2019 {\em Phys.
  Rev.\/} {\bf D100} 044009 (\textit{Preprint} \eprint{1904.13291})

\bibitem{Wette:2018bhc}
Wette K, Walsh S, Prix R and Papa M~A 2018 {\em Phys. Rev.\/} {\bf D97} 123016
  (\textit{Preprint} \eprint{1804.03392})

\bibitem{Keitel:2013wga}
Keitel D, Prix R, Papa M~A, Leaci P and Siddiqi M 2014 {\em Phys. Rev.\/} {\bf
  D89} 064023 (\textit{Preprint} \eprint{1311.5738})

\bibitem{Leaci:2015iuc}
Leaci P 2015 {\em Phys. Scripta\/} {\bf 90} 125001

\bibitem{Suvorova:2016rdc}
Suvorova S, Sun L, Melatos A, Moran W and Evans R~J 2016 {\em \prd\/} {\bf D93}
  123009 (\textit{Preprint} \eprint{1606.02412})

\bibitem{Suvorova:2017dpm}
Suvorova S, Clearwater P, Melatos A, Sun L, Moran W and Evans R~J 2017 {\em
  Phys. Rev.\/} {\bf D96} 102006 (\textit{Preprint} \eprint{1710.07092})

\bibitem{Sun:2017zge}
Sun L, Melatos A, Suvorova S, Moran W and Evans R 2018 {\em \prd\/} {\bf D97}
  043013 (\textit{Preprint} \eprint{1710.00460})

\bibitem{Sun:2018owi}
Sun L and Melatos A 2019 {\em Phys. Rev.\/} {\bf D99} 123003 (\textit{Preprint}
  \eprint{1810.03577})

\bibitem{Viterbi1967}
Viterbi A 1967 {\em IEEE Transactions on Information Theory\/} {\bf 13}
  260--269 ISSN 0018-9448

\bibitem{Bayley:2019bcb}
Bayley J, Woan G and Messenger C 2019 {\em Phys. Rev.\/} {\bf D100} 023006
  (\textit{Preprint} \eprint{1903.12614})

\bibitem{Abbott:2018hgk}
Abbott B~P {\em et~al.\/} (Virgo, LIGO Scientific) 2019 {\em ApJ\/} {\bf 875}
  160 (\textit{Preprint} \eprint{1810.02581})

\bibitem{Keitel:2019zhb}
Keitel D, Woan G, Pitkin M, Schumacher C, Pearlstone B, Riles K, Lyne A~G,
  Palfreyman J, Stappers B and Weltevrede P 2019  (\textit{Preprint}
  \eprint{1907.04717})

\bibitem{Miller:2018rbg}
Miller A {\em et~al.\/} 2018 {\em Phys. Rev.\/} {\bf D98} 102004
  (\textit{Preprint} \eprint{1810.09784})

\bibitem{Oliver:2019ksl}
Oliver M, Keitel D and Sintes A~M 2019 {\em Phys. Rev.\/} {\bf D99} 104067
  (\textit{Preprint} \eprint{1901.01820})

\bibitem{Miller:2019jtp}
Miller A~L {\em et~al.\/} 2019 {\em Phys. Rev.\/} {\bf D100} 062005
  (\textit{Preprint} \eprint{1909.02262})

\bibitem{Meadors:2015vpc}
Meadors G~D, Goetz E and Riles K 2016 {\em Class. Quant. Grav.\/} {\bf 33}
  105017 (\textit{Preprint} \eprint{1512.02105})

\bibitem{Abbott:2019uwg}
Abbott B~P {\em et~al.\/} (LIGO Scientific, Virgo) 2019 {\em Phys. Rev.\/} {\bf
  D100} 122002 (\textit{Preprint} \eprint{1906.12040})

\bibitem{Covas:2020nwy}
Covas P~B and Sintes A~M 2020  (\textit{Preprint} \eprint{2001.08411})

\bibitem{2019RPPh...82a6903C}
{Christensen} N 2019 {\em Reports on Progress in Physics\/} {\bf 82} 016903
  (\textit{Preprint} \eprint{1811.08797})

\bibitem{2018PhRvX...8b1019S}
{Smith} R and {Thrane} E 2018 {\em Physical Review X\/} {\bf 8} 021019
  (\textit{Preprint} \eprint{1712.00688})

\bibitem{gwastro-PENR-RIFT}
{Lange} J, {O'Shaughnessy} R and {Rizzo} M 2018 {\em Submitted to PRD;
  available at arxiv:1805.10457\/}

\bibitem{2010PhRvD..81f2003V}
{Veitch} J and {Vecchio} A 2010 {\em \prd\/} {\bf 81} 062003 (\textit{Preprint}
  \eprint{0911.3820})

\bibitem{Blasco2017}
Blasco A 2017 {\em MCMC\/} (Cham: Springer International Publishing) pp 85--102
  ISBN 978-3-319-54274-4
  \urlprefix\url{https://doi.org/10.1007/978-3-319-54274-4_4}

\bibitem{gwastro-PENR-RIFT-GPU}
{Wysocki} D, {O'Shaughnessy} R, {Lange} J and {Fang} Y~L~L 2019 {\em \prd\/}
  {\bf 99} 084026

\bibitem{2009MNRAS.398.1601F}
{Feroz} F, {Hobson} M~P and {Bridges} M 2009 {\em \mnras\/} {\bf 398}
  1601--1614 (\textit{Preprint} \eprint{0809.3437})

\bibitem{PhysRevLett.122.211101}
Chua A~J~K, Galley C~R and Vallisneri M 2019 {\em Phys. Rev. Lett.\/} {\bf
  122}(21) 211101
  \urlprefix\url{https://link.aps.org/doi/10.1103/PhysRevLett.122.211101}

\bibitem{2020arXiv200207656G}
{Green} S~R, {Simpson} C and {Gair} J 2020 {\em arXiv e-prints\/}
  arXiv:2002.07656 (\textit{Preprint} \eprint{2002.07656})

\bibitem{1705.07057}
Papamakarios G, Pavlakou T and Murray I 2017 Masked autoregressive flow for
  density estimation (\textit{Preprint} \eprint{arXiv:1705.07057})

\bibitem{Singer_2016}
Singer L~P, Chen H~Y, Holz D~E, Farr W~M, Price L~R, Raymond V, Cenko S~B,
  Gehrels N, Cannizzo J, Kasliwal M~M, Nissanke S, Coughlin M, Farr B, Urban
  A~L, Vitale S, Veitch J, Graff P, Berry C~P~L, Mohapatra S and Mandel I 2016
  {\em The Astrophysical Journal\/} {\bf 829} L15
  \urlprefix\url{https://doi.org/10.3847%2F2041-8205%2F829%2F1%2Fl15}

\bibitem{Kapadia:2019uut}
Kapadia S~J {\em et~al.\/} 2019  (\textit{Preprint} \eprint{1903.06881})

\bibitem{Foucart:2012nc}
Foucart F 2012 {\em Phys. Rev.\/} {\bf D86} 124007 (\textit{Preprint}
  \eprint{1207.6304})

\bibitem{Foucart:2018rjc}
Foucart F, Hinderer T and Nissanke S 2018 {\em Phys. Rev.\/} {\bf D98} 081501
  (\textit{Preprint} \eprint{1807.00011})

\bibitem{Chatterjee:2019avs}
Chatterjee D, Ghosh S, Brady P~R, Kapadia S~J, Miller A~L, Nissanke S and
  Pannarale F 2019  (\textit{Preprint} \eprint{1911.00116})

\bibitem{scikit-learn}
Pedregosa F, Varoquaux G, Gramfort A, Michel V, Thirion B, Grisel O, Blondel M,
  Prettenhofer P, Weiss R, Dubourg V, Vanderplas J, Passos A, Cournapeau D,
  Brucher M, Perrot M and Duchesnay E 2011 {\em Journal of Machine Learning
  Research\/} {\bf 12} 2825--2830

\bibitem{0264-9381-27-11-114007}
Mandel I and O'Shaughnessy R 2010 {\em Classical and Quantum Gravity\/} {\bf
  27} 114007 \urlprefix\url{http://stacks.iop.org/0264-9381/27/i=11/a=114007}

\bibitem{2017Natur.548..426F}
{Farr} W~M, {Stevenson} S, {Miller} M~C, {Mandel} I, {Farr} B and {Vecchio} A
  2017 {\em Nature\/} {\bf 548} 426--429 (\textit{Preprint}
  \eprint{1706.01385})

\bibitem{2017MNRAS.471.2801S}
{Stevenson} S, {Berry} C~P~L and {Mandel} I 2017 {\em \mnras\/} {\bf 471}
  2801--2811 (\textit{Preprint} \eprint{1703.06873})

\bibitem{2015ApJ...810...58S}
{Stevenson} S, {Ohme} F and {Fairhurst} S 2015 {\em \apj\/} {\bf 810} 58
  (\textit{Preprint} \eprint{1504.07802})

\bibitem{0264-9381-34-3-03LT01}
Vitale S, Lynch R, Sturani R and Graff P 2017 {\em Classical and Quantum
  Gravity\/} {\bf 34} 03LT01
  \urlprefix\url{http://stacks.iop.org/0264-9381/34/i=3/a=03LT01}

\bibitem{2017PhRvD..95l4046G}
{Gerosa} D and {Berti} E 2017 {\em \prd\/} {\bf 95} 124046 (\textit{Preprint}
  \eprint{1703.06223})

\bibitem{2018arXiv180102699T}
{Talbot} C and {Thrane} E 2018 {\em \apj\/} {\bf 856} 173 (\textit{Preprint}
  \eprint{1801.02699})

\bibitem{2017MNRAS.465.3254M}
{Mandel} I, {Farr} W~M, {Colonna} A, {Stevenson} S, {Ti{\v{n}}o} P and {Veitch}
  J 2017 {\em \mnras\/} {\bf 465} 3254--3260 (\textit{Preprint}
  \eprint{1608.08223})

\bibitem{2017arXiv171202643W}
{Wysocki} D 2017 {\em arXiv e-prints\/} arXiv:1712.02643 (\textit{Preprint}
  \eprint{1712.02643})

\bibitem{2019arXiv190504825P}
{Powell} J, {Stevenson} S, {Mandel} I and {Tino} P 2019 {\em arXiv e-prints\/}
  arXiv:1905.04825 (\textit{Preprint} \eprint{1905.04825})

\bibitem{2020arXiv200209491W}
{Wong} K~W~K, {Contardo} G and {Ho} S 2020 {\em arXiv e-prints\/}
  arXiv:2002.09491 (\textit{Preprint} \eprint{2002.09491})

\bibitem{PhysRevD.100.103025}
Chatterjee C, Wen L, Vinsen K, Kovalam M and Datta A 2019 {\em Phys. Rev. D\/}
  {\bf 100}(10) 103025
  \urlprefix\url{https://link.aps.org/doi/10.1103/PhysRevD.100.103025}

\bibitem{MetzgerBerger2012}
{Metzger} B~D and {Berger} E 2012 {\em \apj\/} {\bf 746} 48 (\textit{Preprint}
  \eprint{1108.6056})

\bibitem{Cowperthwaite2018}
{Cowperthwaite} P~S, {Berger} E, {Rest} A, {Chornock} R, {Scolnic} D~M,
  {Williams} P~K~G, {Fong} W, {Drout} M~R, {Foley} R~J, {Margutti} R, {Lunnan}
  R, {Metzger} B~D and {Quataert} E 2018 {\em \apj\/} {\bf 858} 18
  (\textit{Preprint} \eprint{1710.02144})

\bibitem{Ackley2019}
Ackley K, Eikenberry S~S, Yildirim C, Klimenko S and Garner A 2019  {\bf 158}
  172 \urlprefix\url{https://doi.org/10.3847\%2F1538-3881\%2Fab3c4b}

\bibitem{Flaugher:2015pxc}
Flaugher B {\em et~al.\/} (DES) 2015 {\em Astron. J.\/} {\bf 150} 150
  (\textit{Preprint} \eprint{1504.02900})

\bibitem{Rau:2009yx}
Rau A {\em et~al.\/} 2009 {\em Publ. Astron. Soc. Pac.\/} {\bf 121} 1334--1351
  (\textit{Preprint} \eprint{0906.5355})

\bibitem{Law:2009ys}
Law N {\em et~al.\/} 2009 {\em Publ. Astron. Soc. Pac.\/} {\bf 121} 1395
  (\textit{Preprint} \eprint{0906.5350})

\end{thebibliography}

\end{document}